\documentclass[a4paper,12pt]{article}
\pdfoutput=1
\usepackage{feynmp-auto,expdlist}
\usepackage{amsmath, amsfonts}
\usepackage{graphicx,arydshln}
\usepackage{enumerate}
\usepackage{hyperref}
\usepackage{latexsym}
\usepackage{dsfont}
\usepackage{hepnicenames}
\usepackage{enumerate}
\usepackage{soul}
\usepackage[normalem]{ulem}
\usepackage{comment}
\usepackage[version=4]{mhchem}

\newcommand{\footnoten}[1]{}

\usepackage[font={small},textfont={it}]{caption} 

\usepackage{units}

\usepackage{mathrsfs,graphicx,rotating,amsmath,amsfonts,mathtools,booktabs,amssymb,wasysym}
\usepackage{hyperref}\usepackage{slashed}
\usepackage[nosort]{cite}
\usepackage[table,xcdraw,dvipsnames]{xcolor}
\usepackage{bm}
\usepackage{multirow,multicol}
\hypersetup{colorlinks,bookmarksopen,bookmarksnumbered,
	linkcolor=blus,pdfstartview=FitH,urlcolor=rossos,citecolor=verde}
\allowdisplaybreaks

\newcommand{\UL}{{\cal U}}
\renewcommand{\[}{\left[}

\newcommand{\DM}{{\rm DM}}
\newcommand{\SM}{{\rm SM}}
\newcommand{\GDM }{G_{\rm DM}}
\newcommand{\GF}{G_{\rm F}}

\def\Lag{\mathscr{L}}

\newcommand{\mio}[1]{}

\def\bpm{\begin{pmatrix}}
	\def\epm{\end{pmatrix}}

\usepackage{mathrsfs}

\newcommand{\fig}[1]{~\ref{fig:#1}}
\newcommand{\sfrac}[2]{#1/#2}

\allowdisplaybreaks
\usepackage{multicol}
\usepackage{color}
\definecolor{rosso}{cmyk}{0,1,1,0.4}
\definecolor{rossos}{cmyk}{0,1,1,0.55}
\definecolor{rossoc}{cmyk}{0,1,1,0.2}
\definecolor{blu}{cmyk}{1,1,0,0.3}
\definecolor{blus}{cmyk}{1,1,0,0.6}
\definecolor{bluc}{cmyk}{1,1,0,0.1}
\definecolor{verde}{cmyk}{0.92,0,0.59,0.25}
\definecolor{verdec}{cmyk}{0.92,0,0.59,0.15}
\definecolor{verdes}{cmyk}{0.92,0,0.59,0.4}

\newcommand{\eq}[1]{~{\rm (\ref{eq:#1})}}

\newcommand{\s}{\,{\rm s}}
\newcommand{\keV}{\,{\rm keV}}

\newcommand{\MeV}{\,{\rm MeV}}
\newcommand{\GeV}{\,{\rm GeV}}
\newcommand{\TeV}{\,{\rm TeV}}
\newcommand{\cm}{\,{\rm cm}}

\newcommand{\Tr}{\,{\rm Tr}}

\def\circa#1{\,\raise.3ex\hbox{$#1$\kern-.75em\lower1ex\hbox{$\sim$}}\,}

\newcommand{\beq}{\begin{equation}}
\newcommand{\eeq}{\end{equation}}

\newcommand{\bea}{\begin{eqnarray}}
\newcommand{\eea}{\end{eqnarray}}
\newcommand{\be}{\begin{equation}}
\newcommand{\ee}{\end{equation}}
\font\tenrsfs=rsfs10 at 12pt
\font\sevenrsfs=rsfs7
\font\fiversfs=rsfs5
\newfam\rsfsfam
\textfont\rsfsfam=\tenrsfs
\scriptfont\rsfsfam=\sevenrsfs
\scriptscriptfont\rsfsfam=\fiversfs

\newsavebox\MBox

\renewenvironment{thebibliography}[1]
{\begin{multicols}{2}[\section*{\refname}]%
		\@mkboth{\MakeUppercase\refname}{\MakeUppercase\refname}%
		\list{\@biblabel{\@arabic\c@enumiv}}%
		{\settowidth\labelwidth{\@biblabel{#1}}%
			\leftmargin\labelwidth
			\advance\leftmargin\labelsep
			\@openbib@code
			\usecounter{enumiv}%
			\let\p@enumiv\@empty
			\renewcommand\theenumiv{\@arabic\c@enumiv}}%
		\sloppy
		\clubpenalty4000
		\@clubpenalty \clubpenalty
		\widowpenalty4000%
		\sfcode`\.\@m}
	{\def\@noitemerr
		{\@latex@warning{Empty `thebibliography' environment}}%
		\endlist\end{multicols}}

\newcommand{\eV}{\,{\rm eV}}
\newcommand{\SU}{\,{\rm SU}}

\newcommand{\U}{\,{\rm U}}

\def\circa#1{\,\raise.3ex\hbox{$#1$\kern-.75em\lower1ex\hbox{$\sim$}}\,}
\makeatletter

\font\ital=cmu10

\def\hhref#1{\href{http://arxiv.org/abs/#1}{arXiv:#1}}
\usepackage{xstring}
\newcommand{\hhrefq}[1]{\IfSubStr{#1}{:}{\href{http://inspirehep.net/search?ln=en&ln=en&p=#1&of=hb&action_search=Search&sf=&so=d&rm=&rg=25&sc=0}{InSpire:#1}}{\hhref{#1}}}

\def\art{\@ifnextchar[{\eart}{\oart}}
\def\eart[#1]#2#3#4#5#6{{\rm #2}, {\em #3 \bf #4} {\rm (#6) #5} ({\em #1})}
\def\article{\@ifnextchar[{\earticle}{\oarticle}}
\def\oarticle#1#2#3#4#5#6{{\rm #1}, {\ital ``#6''}, {\rm #2 #3 (#5) #4}}
\def\earticle[#1]#2#3#4#5#6#7{{\rm #2}, {\ital ``#7''}, {\rm #3 #4 (#6) #5}  [\hhrefq{#1}]}
\def\hepart[#1]#2{{\rm #2, \sl#1}}
\def\heparticle[#1]#2#3{#2, {\ital ``#3''} [\hhrefq{#1}]}
\newcommand{\doi}[1]{\href{http://dx.doi.org/#1}{[link]}}

\newcommand{\hhrefqq}[1]{\IfBeginWith{#1}{10.}{\href{https://doi.org/#1}{doi:#1}}{\hhrefq{#1}}}
\def\earticle[#1]#2#3#4#5#6#7{{\rm #2}, {\ital ``#7''}, {\rm #3 #4 (#6) #5}  [\hhrefqq{#1}]}

\renewenvironment{thebibliography}[1]
{\begin{multicols}{2}[\section*{\refname}]%
		\@mkboth{\MakeUppercase\refname}{\MakeUppercase\refname}%
		\list{\@biblabel{\@arabic\c@enumiv}}%
		{\settowidth\labelwidth{\@biblabel{#1}}%
			\leftmargin\labelwidth
			\advance\leftmargin\labelsep
			\@openbib@code
			\usecounter{enumiv}%
			\let\p@enumiv\@empty
			\renewcommand\theenumiv{\@arabic\c@enumiv}}%
		\sloppy
		\clubpenalty4000
		\@clubpenalty \clubpenalty
		\widowpenalty4000%
		\sfcode`\.\@m}
	{\def\@noitemerr
		{\@latex@warning{Empty `thebibliography' environment}}%
		\endlist\end{multicols}}

\newcounter{alphaequation}[equation]
\def\thealphaequation{\theequation\hbox to
	0.6em{\hfil\alph{alphaequation}\hfil}}
\def\eqnsystem#1{
	\def\@eqnnum{{\rm (\thealphaequation)}}
	\def\@@eqncr{\let\@tempa\relax \ifcase\@eqcnt \def\@tempa{& & &} \or
		\def\@tempa{& &}\or \def\@tempa{&}\fi\@tempa
		\if@eqnsw\@eqnnum\refstepcounter{alphaequation}\fi
		\global\@eqnswtrue\global\@eqcnt=0\cr}
	\refstepcounter{equation} \let\@currentlabel\theequation \def\@tempb{#1}
	\ifx\@tempb\empty\else\label{#1}\fi
	\refstepcounter{alphaequation}
	\let\@currentlabel\thealphaequation
	\global\@eqnswtrue\global\@eqcnt=0 \tabskip\@centering\let\\=\@eqncr
	$$\halign to \displaywidth\bgroup \@eqnsel\hskip\@centering
	$\displaystyle\tabskip\z@{##}$&\global\@eqcnt\@ne
	\hskip2\arraycolsep\hfil${##}$\hfil& \global\@eqcnt\tw@\hskip2\arraycolsep
	$\displaystyle\tabskip\z@{##}$\hfil
	\tabskip\@centering&\llap{##}\tabskip\z@\cr}
\def\endeqnsystem{\@@eqncr\egroup$$\global\@ignoretrue} \makeatother

\definecolor{Gray}{gray}{0.95}

\def\bal#1\eal{\begin{align}#1\end{align}}

\newcommand{\GDMT}{\Gamma_{{\rm DM}T}}

\begin{document}
\thispagestyle{empty}

\begin{center}  
{\huge\bf\color{rossos}Dark Matter abundance via thermal \\[1ex] decays and leptoquark mediators}
\vspace{1cm}

{\bf Benedetta Belfatto$^{a,b}$, Dario Buttazzo$^a$, Christian Gross$^b$,\\ Paolo Panci$^{a,b}$, Alessandro Strumia$^b$, Natascia Vignaroli$^{c,d}$,\\ Ludovico Vittorio$^{e,a}$, Ryoutaro Watanabe$^a$}  \\[6mm]
{\it $^a$ INFN Sezione di Pisa, Italia}\\[1mm]
{\it $^b$ Dipartimento di Fisica, Universit\`a di Pisa, Italia}\\[1mm]
{\it $^c$ Dipartimento di Fisica, Universit\`a di Napoli ``Federico II", Italia}\\[1mm]
{\it $^d$ INFN Sezione di Napoli, Italia}\\[1mm]
{\it $^e$ Scuola Normale Superiore, Pisa, Italia}\\[1mm]

\vspace{1cm}

{\large\bf Abstract}

\begin{quote}\large
We explore a new mechanism for reproducing the Dark Matter (DM) abundance: 
scatterings of one DM particle on light Standard Model particles. 
Strong bounds on its decays can be satisfied if DM undergoes freeze-in and has a mass around or below the pion mass.  
This happens, for example, in theories with a right-handed neutrino interacting with charged fermions through a leptoquark exchange.
These leptoquarks can be linked to the ones motivated by the $B$-physics anomalies if assumptions about the flavour structure are made.
DM signals are unusual, with interesting possibilities for direct and indirect detection.
Achieving thermal freeze-out instead requires models with more than one DM flavour, and couplings parametrically smaller than what needed by the usual pair annihilations.
\end{quote}
\end{center}

\setcounter{page}{1}
\newpage
\tableofcontents	
\section{Introduction}
The measured value of the Dark Matter (DM) energy density can be understood in terms of
various hypothetical scattering processes occurring in the early universe.
The most studied scenario considers $2\leftrightarrow 0$ annihilation processes, where we count the number of 
DM-sector particles plus anti-particles
in the initial and final state, without indicating the Standard Model (SM) particles. 
Examples of such processes that change the number of DM (anti)particles by two units are
\beq \label{eq:DMann} {\rm DM}\,  {\rm DM} \leftrightarrow {\rm SM}\, {\rm SM},\qquad
{\rm DM}\,  {\rm DM} \leftrightarrow {\rm SM}\, {\rm SM}\, {\rm SM} . \eeq
A $\mathbb{Z}_2$ symmetry acting on the dark sector only usually justifies the
presence of 2 DM (or, more generically, dark sector) particles and implies DM stability.
These annihilations (or, more generically, co-annihilations) can lead to DM around the TeV scale
as a freeze-out thermal relic. A freeze-in non-thermal relic and other possibilities can also arise through the same interactions.

\smallskip

Processes more complicated than $2\leftrightarrow 0$ received less attention:
$2\leftrightarrow 1$ scatterings, known as semi-annihilations, can arise in models where DM is stable, and behave similarly to annihilations (see~\cite{0811.0172,1003.5912,2103.16572} for recent references).
$3\leftrightarrow 2$ or $4\leftrightarrow 2$ scatterings give qualitatively different behaviours sometimes
dubbed `cannibalistic DM'~\cite{Cannibalism} or `pandemic DM'~\cite{2103.16572}, depending on the regime
and on whether DM is  in thermal equilibrium with the SM sector.
These scatterings can reproduce the DM cosmological density for DM masses in the eV or MeV range,
and can be justified by $\mathbb{Z}_3$ or higher symmetries.

Little attention has been received by the simplest possibility: $1\leftrightarrow0$ processes
that could be dubbed `decadent DM'.
Examples of such transitions that change the DM number by one unit are
 \beq  \label{eq:thermaldec}
 {\rm DM}\,  {\rm SM} \leftrightarrow {\rm SM}\, {\rm SM},\qquad
  {\rm DM} \leftrightarrow {\rm SM}\, {\rm SM},\qquad
   {\rm SM} \leftrightarrow {\rm DM}\, {\rm SM}.\eeq 
These processes arise in absence of a symmetry that forbids couplings between
one DM particle and the SM sector. 
%
This is what can happen, for instance, in presence of interactions under which both DM and SM fields are charged, where
imposing a symmetry that separates the dark sector from the SM might be not be a viable option.
Prototypical examples of such a scenario include ``sterile neutrino'' DM, that can couple to leptons via a Higgs boson or a right-handed vector boson, or models where DM interacts with quarks via a leptoquark mediator. In both cases, the mediator carries a conserved SM charge and is allowed to couple to a DM-SM current.

The latter case of DM interactions mediated by a leptoquark is particularly interesting in the context of models that aim at explaining the anomalies observed in $B$-meson decays (see e.g.\ \cite{Buttazzo:2017ixm,Azatov:2018kzb,Isidori:2021vtc,Cornella:2021sby,Fajfer:2015ycq,Bauer:2015knc,Barbieri:2017tuq,DiLuzio:2017vat} and references therein).
The possibility of DM coupled to leptoquarks has been considered before~\cite{Mandal:2018czf,Baker:2021llj,Belanger:2021smw,Guadagnoli:2020tlx,Baker:2015qna}, but usually in combination with an additional $\mathbb{Z}_2$ symmetry that forbids DM decay. Here we shall explore the minimal possibility that allows the mediator to simultaneously couple to DM-SM and pure SM currents.

Given that $1\leftrightarrow0$ processes are unavoidable in some theories, we show that,
on the other hand, they can play a useful role in cosmology.  
Indeed, since DM needs to scatter on SM particles which are abundant in the primordial plasma,
in order to reproduce the DM relic abundance these processes need a smaller cross section than that of
the usual DM annihilations in eq.\eq{DMann}, 
where DM needs instead to find a rare DM particle.
We refer to the interactions in eq.\eq{thermaldec} as {\em``thermal decays''} because
 at finite temperature such transitions are conveniently described by Boltzmann equations with
a thermal width $\Gamma_T$ of DM.
This terminology emphasises the obvious and potentially insurmountable problem:
DM risks decaying too fast, since there is no separate dark sector with a conserved DM number.
Interactions that lead to the thermal decays in eq.\eq{thermaldec}
also lead to DM decays.
Let us be quantitative:
\begin{itemize}
\item On one hand,
the DM life-time $\tau_{\rm DM} = 1/\Gamma_{\rm DM}$ must surely be longer than 
the age of the Universe $\tau_{\rm DM} >T_U (\sim 10^{18}\,{\rm s})$. 
Many DM decay modes are more strongly constrained, requiring
$ \tau_{\rm DM} \circa{>} 10^{26}\,{\rm s}$~\cite{1309.4091,1612.07698,2008.01084}.

\item On the other hand, thermal decays can give non-relativistic DM freeze-out\footnote{Freeze-in needs
a smaller $\GDMT$. We do not consider relativistic DM
freeze-out (analogous to neutrino decoupling), because the resulting DM abundance
would be of order unity
$Y_{\rm DM}\equiv n_{\rm DM}/s =45\zeta(3)g_{\rm DM}/g_{\rm SM}(T_{\rm dec})\pi^4$
and thereby the DM cosmological abundance would be reproduced for
 warm DM with mass $M \approx 0.7 \eV\, g_{\rm SM}(T_{\rm dec})/g_{\rm DM}$.
To avoid problematic warm DM, 
the DM decoupling temperature $T_{\rm dec}$ must be mildly below the DM mass $M$.
So we need a sizeable $\Gamma_{{\rm DM}T} \sim H$ at $T \sim M$.}
reproducing
the cosmological DM abundance if their rate $\GDMT$ is comparable
to the Hubble rate $H \sim \sqrt{d_{\rm SM}} (T/\GeV)^2/(5~10^{-6}{\rm s})$ 
at temperatures around the DM mass $M$, where $d_{\rm SM}$ is the number of
SM degrees of freedom.

\end{itemize}
As a result, the two requirements conflict easily and by many orders of magnitude: one needs
\beq \label{eq:ahi}
\Gamma_{\rm DM}/\Gamma_{{\rm DM}T} \circa{<}10^{-30}.\eeq
This gap by 30 orders of magnitude is such a strong constraint on possible models
that this scenario is widely ignored. 
We nevertheless explore this possibility: while the simplest freeze-out picture in models with one DM particle is clearly ruled out, more complex possibilities with multiple DM particles or with DM freeze-in 
emerge and can lead to a peculiar phenomenology.
A common feature of all viable models is the presence of some small coupling between DM and SM, which could be related to an underlying flavour structure.


\smallskip
The paper is organized as follows.
In section~\ref{Boltzmann} we write the generic form
of Boltzmann equations
that describe DM with thermal decays and study their features, deriving approximate solutions.
We find that the DM relic abundance depends on a different power of the cross-section than usual DM annihilations.
As a result DM thermal decays can be more efficient than  DM annihilations,
and can match the cosmological DM abundance with rates not much stronger
than the minimal ones needed to reach thermal equilibrium.
DM freeze-in needs smaller sub-equilibrium rates.

In section~\ref{DM1} we discuss possible models that only involve one DM particle.
DM has to be lighter than hundreds of MeV in order to close the phase space for tree-level decays into SM particles.
We then focus on models where DM carries lepton number (e.g.\ a right-handed neutrino) and interacts with the SM via TeV-scale leptoquarks motivated by the flavour anomalies.
%
We will find that such models can reproduce the DM abundance via freeze-in, but not via freeze-out.

In section 4 we study possible signatures. We first explore possible values of the DM couplings motivated by simple flavour patterns, and the possible relations with $B$-physics anomalies.
We then discuss direct and indirect detection.
Interactions involving one DM particle give the unusual direct detection signals discussed in section~\ref{DD},
suggesting a tentative explanation of the {\sc Xenon1T} anomaly.
Indirect detection signals are discussed in section~\ref{ID}, where we show that 
DM decays $\DM \to \nu\gamma$ could explain the line claimed at $3.5\keV$,
compatibly with cosmological production via freeze-in of thermal decays. 

Finally, in section~\ref{DM2} we show that freeze-out is compatible with DM stability in models
with two or more DM particles (for example three generations of  right-handed neutrinos).
Conclusions are given in section~\ref{concl}.



\section{Boltzmann equation for thermal decays}\label{Boltzmann}
As usual, it is convenient to write the Boltzmann equation
for $Y\equiv n/s$ in terms of $z=M/T$, where $M$ is the DM mass,
$n$ is the DM number density, and 
$s$ the total entropy.
Including scatterings that involve two DM
(the usual DM annihilations) and one DM
(our thermal decays) the Boltzmann equation is
\beq \label{eq:DMboltzY}
sHZz \frac{dY}{dz} =
  -2\gamma^{\rm eq}_{\rm ann}   \bigg(\frac{Y^2}{Y^{2}_{\rm eq}}-1\bigg) 
    -\gamma^{\rm eq}_{\rm dec}   \bigg(\frac{Y}{Y_{\rm eq}}-1\bigg) .
 \eeq
where the factor $ Z=1/(1+ \frac{1}{3}\frac{d\ln d_{\rm SM}}{d\ln T})$ can often be approximated as 1.
In order to match the DM cosmological density, DM must freeze-out while non-relativistic,
so that its thermal equilibrium abundance is Boltzmann suppressed,
$n_i^{\rm eq} \simeq d_i(m_iT/2\pi)^{3/2} e^{-m_i/T}$,
where $d_i$ is the number of degrees of freedom for the particle $i$.
The space-time density of scatterings $\gamma_i^{\rm eq}$
can thereby be simplified going to the non-relativistic limit.
The initial state of the usual
${\rm DM}\,  {\rm DM} \leftrightarrow {\rm SM}\, {\rm SM}$ annihilations becomes non-relativistic at low temperature.
When considering more generic processes an unusual step is needed:
identifying which side of the reaction becomes non-relativistic,
namely the side that involves heavier particles.
Defining as $A,B$ the particles that can scatter at rest
(they can be in the initial state or in the final state, depending on the kinematics of each process),
the rate is approximated as
\beq \gamma^{\rm eq}_i  \stackrel{T\ll m}{\simeq}   S n_A^{\rm eq} n_B^{\rm eq}  \sigma_{i0}\qquad
\hbox{where}\qquad
 \sigma_{i0}= \lim_{v\to 0}\langle \sigma_i v\rangle. \eeq
The initial-state symmetry factor $S$ is only present when the non-relativistic
particles are in the initial state.
In the presence of multiple processes, the total $\gamma_i^{\rm eq}$ are computed by
summing their rates.
In the non-relativistic limit the Boltzmann equation becomes
\beq \label{eq:DMboltzY}
 \frac{dY}{dz} =
   -f_{\rm ann}(z) (Y^2-Y^{2}_{\rm eq}) -f_{\rm dec}(z)  {\cal B}(z)
   (Y-Y_{\rm eq})
 \eeq
where we defined \beq
  f_i (z)\equiv \frac{ s\sigma_{0i}}{H Zz}
{\simeq} \frac{\lambda_i}{z^2}\,,\qquad
 \lambda_i =
M_{\rm Pl} M \sigma_{0i}\sqrt{\frac{\pi d_{\rm SM}}{45}}  \,,
  \eeq
such that DM is in thermal equilibrium at $T \sim M$ if $\lambda_i \gg 1$. 
In the annihilation case, $\lambda_{\rm ann}$ is the usual dimension-less combination
that helps obtaining approximate solutions in the limit $\lambda_{\rm ann} \gg 1$. 
The thermal scattering term contains an extra unusual but important factor $B(z)$,
since it contains a Boltzmann suppression.
Its non-relativistic approximation is
\beq  
 {\cal B} \equiv \frac{Y^{\rm eq}_{A}Y^{\rm eq}_{B}}{Y^{\rm eq}}\simeq
 {\cal R} \frac{d_{\rm DM}}{s}\left(\frac{MT}{2\pi}\right)^{3/2}\exp\left[- \frac{\Delta}{T}\right] \eeq
where
\begin{align}
\mathcal{R} &= \frac{d_A d_B}{d_{\rm DM}^2} \left(\frac{m_A m_B}{M^2}\right)^{3/2}, & \Delta &= m_A + m_B - M, & &\text{heavier final state}\\
\mathcal{R} &= \frac{d_B}{d_{\rm DM}}\left(\frac{m_B }{M}\right)^{3/2}, & \Delta &= m_B. & &\text{heavier initial state}
\end{align}
The $\Delta$ factor that controls the exponential suppression is a combination of the masses
$M$ of DM, and $m_A$ and $m_B$ of the particles that become non-relativistic.
The important observation is that, in general, $\Delta \neq M$.
The DM mass $M$ is the only relevant mass that controls how
Boltzmann-suppressed the usual DM annihilations into light SM particles are.

One has $m_A = M$ and thereby $\Delta=m_B$ for processes
where the initial state is heavier and hence non-relativistic.
As we are considering ${\rm DM}_A\,  {\rm SM}_B \leftrightarrow {\rm SM}\, {\rm SM}$
processes, $\Delta$ is the mass of the initial-state SM particle.
If instead the final-state is heavier and hence non-relativistic,
${\rm DM}\,  {\rm SM} \leftrightarrow {\rm SM}_A\, {\rm SM}_B$, 
$\Delta$ is given by a  more complicated combination of masses. 

In the presence of multiple process, the one with smallest $\Delta$, 
i.e.\ least Boltzmann suppression, tends to dominate.
The explicit expressions for $\Delta$ imply 
that the Boltzmann suppression is only avoided at the kinematical border where DM 
decay starts being kinematically allowed.

We see that the `thermal decay' terms corresponds to 
an effective decay width  $\Gamma_{{\rm DM}T} = B \sigma_{\rm dec0}$ that gets Boltzmann suppressed at low $T$. 
Indeed the rate of a true decay with width $\Gamma_{{\rm DM}T}$ would be
$\gamma_{\rm dec}= \Gamma_{{\rm DM}T} n_1^{\rm eq}
{\mbox{K}_1(z)/\mbox{K}_2(z)}\, \stackrel{z\gg 1}\simeq  n_1^{\rm eq}\Gamma_{{\rm DM}T}$.

\begin{figure}[t]
$$ \includegraphics[height=0.35\textwidth]{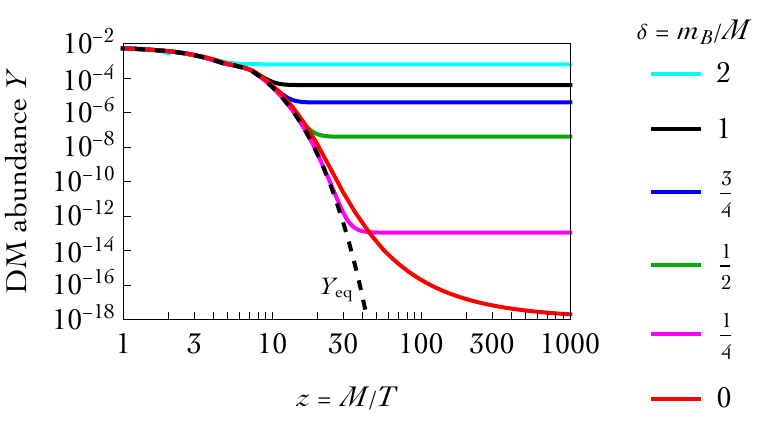}
$$
\caption{Evolution of the DM abundance $Y=n/s$ for thermal decays,
assuming a heavier initial-state involving a SM particle with mass $m_B = \delta\, M$
and DM mass $M=1\GeV$.
The case $\delta=1$ corresponds to DM annihilations.
For $\delta \neq 0$ we considered scatterings with heavier initial state, 
interaction strength $\lambda_{\rm dec}=10^6$,
degrees of freedom $d_{\rm DM}=2$ and $d_{\rm B}=4$.
For $\delta=0$ (red curve) we instead assumed $d_{\rm B}=2$ and a smaller
$\lambda_{\rm dec}=10^{4}$: despite this it produces a lower DM relic abundance. 
\label{fig:Yevo}}
\end{figure}

\subsection{Approximate solution to the Boltzmann equations: freeze-out}\label{freezeout}
The freeze-out abundance produced by DM annihilations or by thermal decays can be approximatively computed
as follows. These processes reach thermal equilibrium if $\lambda_{\rm ann,dec}\gg1$.
Long before freeze-out, i.e.,\ at early
$z\ll z_f$, we can expand eq.\eq{DMboltzY} to first order in small $Y-Y_{\rm eq}$, finding for $\lambda_i\gg 1$
\beq 
Y(z) -Y_{\rm eq}  \stackrel{z\ll z_f}{\simeq}   - \frac{z^2 Y'_{{\rm eq}}}{2\lambda_{\rm ann} Y_{\rm eq}+\lambda_{\rm dec}  {\cal B}}=
 \frac{z(z-3/2)   }{2 \lambda_{\rm ann} +  \lambda_{\rm dec}  {\cal R}  e^{z(1-\delta)}}.
\label{eq:beforefreezeout}
\eeq
where $\delta \equiv \Delta/M$ so that ${\Delta /T} =z\delta$.
We see that $\delta<1$ i.e.\ $\Delta <M$
enhances the effect of thermal decays compared to thermal annihilations.
In general, the values of $\lambda_{\rm dec}/\lambda_{\rm ann}$ and of $\delta$ decide 
whether freeze-out is determined by annihilations or decays.

The final abundance is roughly approximated as $Y_\infty \sim Y_{\rm eq}(z_f)$ where
the freeze-out temperature $z_f$ is estimated imposing $Y- Y_{\rm eq}=  c Y_{\rm eq}$
with $c \approx 1$.
In the usual case where only DM annihilations are present, $\lambda_{\rm dec}=0$,
one gets the usual estimate for the freeze-out temperature
$z_f \approx \ln \left[2 \lambda_{\rm ann}/(d_{\rm SM} \sqrt{z_f})\right]$, and the usual freeze-out  abundance
$Y_\infty \sim z_f^2/2\lambda_{\rm ann}$.

\begin{itemize}

\item If instead only DM thermal decays are present,  $\lambda_{\rm ann} = 0$,
one gets
\beq \label{eq:zf}
z_f \approx \frac{1}{2\delta}W\left[\frac{ \delta}{\pi}   \left(\frac{45  c \lambda_{\rm dec} {\cal R}  d_{\rm DM}}{4 \pi^3 d_{\rm SM} }\right)^2\right] , 
\eeq
where $\delta = \Delta/M$ and $W(x)$ is the `product log' function, that solves $x=W e^W$
and is roughly approximated by $\ln x$ as $x\to\infty$.
The final DM abundance is approximated as\footnote{This expression is accurate up to order unity factors,
as we neglected  evolution at $z > z_f$.
A more precise expression is obtained matching at the appropriate $z_f$
the approximation of eq.\eq{beforefreezeout} valid at $z\ll z_f$
with the approximation at $z\gg z_f$,
$Y_\infty = Y(z_f) \exp(-\lambda_{\rm dec} \int_{z_f}^\infty dz\,{\cal B}(z)/z^2)$.
However the matching does not result in a simple analytic expression.}
\beq \label{eq:Yinfdec}
Y_\infty \approx  Y_{\rm eq}(z_f) \sim  \lambda_{\rm dec}^{-1/\delta} , 
\eeq
showing that thermal decays need a smaller value of $\lambda \gg1$ if $\delta <1$, i.e.\ $\Delta<M$.
Fig.\fig{Yevo} shows examples of numerical solutions to the Boltzmann equations for fixed interaction strength
$\lambda_{\rm dec}$
and varying $\delta$: we see that smaller $\delta$ leads to a lower DM final abundance in agreement with eq.\eq{Yinfdec}.

\item  Thermal decays are maximally efficient for $\delta \to 0$. Let us discuss if/how $\delta=0$ can be realised. 
A look at kinematics shows that this limit corresponds to processes where DM scatters with 
a much lighter SM particle $B$, such as the photon.
Thereby this case only arises near to the dangerous kinematical border where 
the phase space for DM decays starts opening up.
Cancellation of IR divergences between real and virtual effects starts being relevant.
In this special case $ {\cal B}=Y_B^{\rm eq}$ is constant
(so that the non-relativistic limit used in previous approximation no longer holds);
decoupling happens at $z_f \approx \sqrt{\lambda_{\rm dec} Y_B^{\rm eq}}$ (no longer log-suppressed);
subsequent evolution is non-negligible leading to
\beq Y_\infty\approx Y_{\rm eq}(z_f) e^{-\lambda_{\rm dec} Y_B^{\rm eq}/z_f} \approx
\frac{45 d_{\rm DM} (\lambda_{\rm dec} Y_B^{\rm eq})^{3/4}}{2^{5/2}\pi^{7/2} d_{\rm SM}} e^{-2\sqrt{\lambda_{\rm dec} Y_B^{\rm eq}}}.\eeq
This case is also illustrated in fig.\fig{Yevo}.
\end{itemize}

\medskip


Let us next discuss what these results mean in practice.
We can approximate a non-relativistic cross section at $T \sim M$ as 
\beq \label{eq:DMcrosssection}
\sigma_0 \approx g_{\rm SM}^p  \left\{ \begin{array}{ll}
g_{\rm DM}^2/4\pi M^2 & \hbox{dimension-4 interaction with coupling $g_{\rm DM}$}\\
\GDM ^2 M^2/4\pi  & \hbox{dimension-6 interaction with coupling $G_{\rm DM}$}
\end{array}\right.\eeq
where the interaction of DM with the SM sector is described by some dimension-4 interaction
with dimension-less coupling $g_{\rm DM}$, or by some dimension-6 operator
with coefficient $\GDM $ with mass dimension $-2$.
Depending on how many particles are involved in this DM interaction, 
extra couplings can be needed to obtain a cross-section, 
so that we add
extra powers $p$ of a generic SM coupling (such as $e$ or $g_2$ or $f_\pi/m_\pi$).
Since its power is model-dependent, for the purpose of the present parametrisation we assume $g_{\rm SM} \sim 1$.

The condition $\lambda\circa{>} \lambda_{\rm min}$ translates into
\beq \label{eq:thermaleq}
g_{\rm DM} \circa{>} 10^{-9}  \sqrt{\frac{ \lambda_{\rm min} M}{\GeV}}, \qquad 
\GDM \circa{>} 10^{-4}\GF \sqrt{\lambda_{\rm min}}\left(\frac{\GeV}{M}\right)^{3/2} 
\qquad\hbox{(thermal equilibrium})\eeq
where $\GF$ is the Fermi constant.
 Both annihilations and thermal decays need at least to reach thermal equilibrium,
corresponding to $\lambda_{\rm min}\sim 1$.
DM annihilations give the freeze-out abundance $Y_\infty \sim 1/\lambda_{\rm ann}$,
so  the cosmological DM abundance is reproduced for $\lambda \sim M/\eV \gg1$,
corresponding to
\beq \label{eq:gthermal}
g_{\rm DM}\sim 10^{-4} \frac{M}{\GeV},\qquad \GDM \sim 10\, \GF \frac{\GeV}{M}\qquad
\hbox{(DM annihilations)}.\eeq
For $\delta <1$, DM thermal decays need DM couplings smaller than in eq.\eq{gthermal},
possibly down to the minimal value in eq.\eq{thermaleq} for $\lambda_{\rm min}\sim 1$,
reached in the limit $\delta \ll 1$ i.e.\ $\Delta \ll M$ where DM thermal decays become maximally efficient.

\begin{figure}[t]
\centering%
\includegraphics[height=0.42\textwidth]{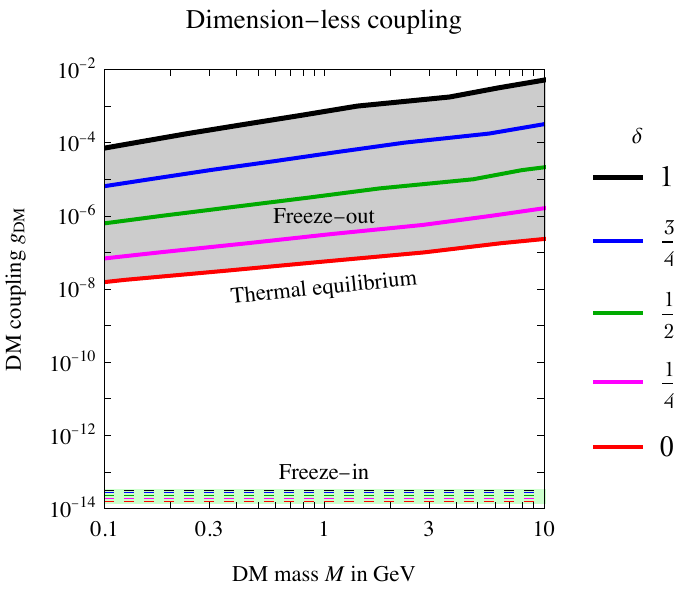}\hfill%
\includegraphics[height=0.42\textwidth]{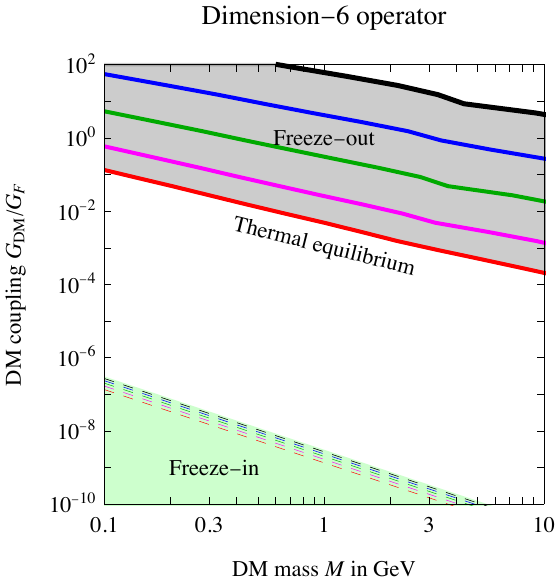}
\caption{Values of the DM coupling to the SM needed to reproduce the DM cosmological abundance
via thermal decays with exponent $\delta =\Delta/M$.
The case $\delta=1$ reproduces DM annihilations.
We assume the cross section in eq.\eq{DMcrosssection} with coupling $g_{\rm SM}=1$ and
other specifications as in fig.\fig{Yevo}. 
{\bfseries Left:} DM with a dimension-less coupling $g_{DM}$ to the SM.
{\bfseries Right:} DM coupled via a dimension-6 operator with coefficient $\GDM$.
Its freeze-in estimate is the minimal contribution that arises if $T_{\rm RH}\sim M$.
\label{fig:NeededgG}}
\end{figure}

Fig.\fig{NeededgG} shows the values of
$g_{\rm DM}$ and $\GDM$ needed to reproduce the DM relic abundance via freeze-out of thermal decays,
assuming $g_{\rm SM}=1$, 
considering different values of $0\le \delta \le 1$ (solid curves).
The needed values range between those characteristic freeze-out of DM annihilations
($\delta=1$, eq.\eq{gthermal} and black curves in fig.\fig{NeededgG})
down to those needed to achieve thermal equilibrium
($\delta=0$, eq.\eq{thermaleq} and red curves in fig.\fig{NeededgG}).
For $\delta <1$ the maximal DM mass $M$ allowed by unitarity is above $100\TeV$
(the upper limit in case of DM annihilations): but model-building considerations will point to much lighter sub-GeV DM with very small couplings.

\subsection{Approximate solution to the Boltzmann equations: freeze-in}\label{freezein}
If DM couplings are so small that it never thermalized, one can instead assume
a vanishing initial DM abundance (needless to say, this is a more questionable assumption than thermal equilibrium)
and match the small amount of generated DM with the observed DM abundance.

The final DM abundance produced by freeze-in is obtained by setting $Y=0$ in the right-handed side of eq.\eq{DMboltzY}. 
One gets 
\beq \label{eq:FIestimate}
Y_\infty = \int_0^\infty \frac{dz}{z}   \frac{\gamma^{\rm eq}_{\rm dec} }{H s}
\sim  \max_T \frac{\gamma^{\rm eq}_{\rm dec}}{H s} \sim \max_T \frac{\Gamma_{{\rm DM}T}}{H}.\eeq
The  DM abundance suggested by cosmology, $Y_\infty \sim \eV/M$, is thereby reproduced 
for a comparably small value of $\Gamma_{{\rm DM}T}/H$.
Unlike in the freeze-out case, replacing DM annihilations with DM thermal decays does not lead to significant changes in the dynamics.
What changes is that the needed scattering can be partially mediated by SM couplings, that can have significant values $g_{\rm SM}\sim 1$. 
Freeze-in can be dominated at low $T \sim M$ or at high $T \gg M$ depending on the
dimension of the interaction:
\begin{itemize}
\item If DM has a dimensionless coupling $g_{\rm DM}$, 
the rate $\gamma^{\rm eq}_{\rm dec}  \sim  g_{\rm SM}^2 g_{\rm DM}^2 T^4$ at $T \gg M$ gives
a DM abundance $Y_\infty$ dominated at $T\sim M$, 
$Y_\infty \sim g_{\rm SM}^2 g_{\rm DM}^2 M_{\rm Pl}/M$.
This matches the cosmological DM abundance for
\beq \label{eq:gDMfin}
g_{\rm DM} g_{\rm SM} \sim {\sqrt{T_0/M_{\rm Pl}}} \sim 10^{-15}\qquad \hbox{(freeze-in of thermal decays)}.\eeq
Precise order one factors depend on the process (such as a scattering or a decay, see~\cite{0911.1120} for examples).
A general expression, valid up to these process-dependent order one factors, is obtained by inserting in eq.\eq{FIestimate} the universal non-relativistic limit 
\beq \label{eq:FIapprox}
Y_\infty \approx \int_0^\infty dz \, f_{\rm dec} B Y_{\rm eq} =\lambda_{\rm dec} \frac{2025 d_{\rm DM}^2 R_B}{32\pi^7 d_{\rm SM}^2 
(1+\delta)^2}.\eeq
This is plotted in the left panel of fig.\fig{NeededgG} assuming the cross section in eq.\eq{DMcrosssection} with $g_{\rm SM}\sim1$. 

\item If DM has a non-renormalizable interaction $\GDM$
one has $\gamma^{\rm eq}_{\rm dec} \sim g_{\rm SM}^2 \GDM^2 T^8$ so that DM production is dominated at $T \gg M$:
the result depends on the full model.
For example, the non-renormalizable operator could be produced by a mediator with mass $M_{\rm med}$ and renormalizable couplings.
If the reheating temperature satisfies $T_{\rm RH} \gg M_{\rm med}$ the DM abundance is dominated at $T \sim M_{\rm med}$
and can be approximated in a way similar to eq.\eq{FIestimate}.  
If $M \ll T_{\rm RH}\ll  M_{\rm med}$ the DM abundance is dominated at $T \sim T_{\rm RH}$,
$Y_\infty \sim  g_{\rm SM}^2 \GDM^2 M_{\rm Pl} T^3_{\rm RH}$.
Eq.\eq{FIestimate} provides the minimal contribution that arises at $T\sim M$ if $T_{\rm RH}\sim M$, 
plotted as upper bound in the right panel of fig.\fig{NeededgG}.  
\end{itemize}

Finally, electro-weak symmetry breaking can dominantly contribute to
the masses of some particles. While this phenomenon can enhance freeze-in
above the weak scale (for example, making the relevant particles all nearly massless),
 it cannot provide a qualitatively new way
of getting long-lived DM with desired cosmological abundance.

\begin{table}[t]
$$\begin{array}{ccccc}
\hbox{particle} & \hbox{life-time} & \hbox{mass} &\epsilon= \Gamma/m & \hbox{main decays}\\ \hline
e & \infty & 0.511\MeV & 0 & \hbox{stable}\\
p =duu& \circa{>} 10^{34}\,{\rm yr} & 938.3\MeV & \circa{<}10^{-66} & \hbox{stable?}\\
n =ddu & 880\s & 939.6\MeV & 0.8~10^{-27} & p e\bar\nu\\
\mu & 2.2\,10^{-6}\s & 105.7\MeV & 2.8~10^{-18} & e \nu_\mu\bar\nu_e\\
K^- = s\bar u& 1.2\,10^{-8}\s& 493.7\MeV & 1.1~10^{-16} & \mu \bar\nu_\mu ,\pi^-\pi^0 \\
\pi^- = d\bar u& 2.6\,10^{-8}\s& 139.6\MeV & 1.8~10^{-16} & \mu \bar\nu_\mu 
\\
\Omega^-  =sss & 0.8 \,10^{-10} \s& 1672\MeV &4.8\, 10^{-15}&\Lambda K^-,\Xi \pi\\
\Sigma^+ = uus & 0.8\,10^{-10}\s & 1189\MeV &6.9\, 10^{-15}&  p\pi^0, n\pi^+\\ 
\end{array}$$
\caption{\label{tab:narrow}SM particles with lowest narrowness $\epsilon = \Gamma/m$.
One component out of multiplets is selected.}
\end{table}%

\section{DM with lepton number and freeze-in}\label{DM1}
We seek theories where DM number is significantly violated, but where
DM is very long lived.
We consider DM coupled to two (or more) SM particles with masses $m_1$ and $m_2$:
\beq  \label{eq:DM12}
{\rm DM} ~{\rm SM}_1~ {\rm SM}_2.\eeq
To avoid the most immediate danger, the tree-level decay
$ {\rm DM} \to {\rm SM}_1\, {\rm SM}_2$,
we assume $M< m_1 + m_2$ so that this process is kinematically blocked.
We also need to assume $M \sim m_{1,2}$, so that
the SM particles take active part in DM decoupling.
Finally, if the coupling \eqref{eq:DM12} is large, it might be necessary to assume $M >|m_1 - m_2|$, to avoid
potentially problematic decays of SM particles into DM.

\bigskip

The next danger is virtual decays such as $ {\rm DM} \to {\rm SM}_1\, {\rm SM}_2^*$.
The $\SM_2$ particle with mass $m_2$ and $d_2$ degrees of freedom is off-shell:
the virtual $\SM_2^*$ has quadri-momentum $k_2$ with $x \equiv k_2^2/m_2^2<1$
and decays into lighter SM particles $\SM_{3,4,\ldots}$ not directly coupled to DM.
The $\DM$ decay width averaged over polarizations is
\beq \Gamma ( \DM  \to \SM_1 \SM_{3,4\ldots}) =
 \int \frac{dx\sqrt{x}}{(x-1)^2}\,  \Gamma(\DM \to \SM_1 \SM_2^*)   \frac{\Gamma(\SM_2^*\to \SM_{3,4,\ldots})}{d_2  \pi m_2} . \eeq
The integral is dominated by the highest available $x_{\rm max}<1$, so that
some decay modes of $\SM_2$ might be closed for $\SM_2^*$.
The off-shell decay width is thereby approximated as
 \beq \Gamma ( \DM  \to \SM_1 \SM_{3,4\ldots}) \approx \frac{\epsilon_2}{4\pi }
  \frac{  \Gamma ( \DM  \to \SM_1\SM_2^*)}{1-x_{\rm max}},\qquad
  \epsilon_i  \equiv \hbox{BR}_i \frac{\Gamma_i}{m_i},\eeq
where $\Gamma_i$ is the total decay width of ${\rm SM}_i$,
and ${\rm BR}_i$ (possibly equal to 1) its branching ratios into light enough particles
(such as photons, neutrinos, electrons).
Each SM$_i$ particle has a value of $\epsilon_i$,
and its smallness is a figure-of-merit for coupling long-lived DM to
SM particles.
Table~\ref{tab:narrow} lists the most narrow SM particles. All of them are lighter than about a GeV
(heavier weak-scale SM particles decay too fast for our purposes),
so DM too has to be similarly light.
At sub-GeV energy, physics is described by baryons and mesons, rather than by quarks.
In order to possibly reproduce the DM abundance via freeze-out, DM interactions
cannot be suppressed by a scale much above the weak scale.

\begin{figure}[t]
\centering%
\includegraphics[width=0.4\textwidth]{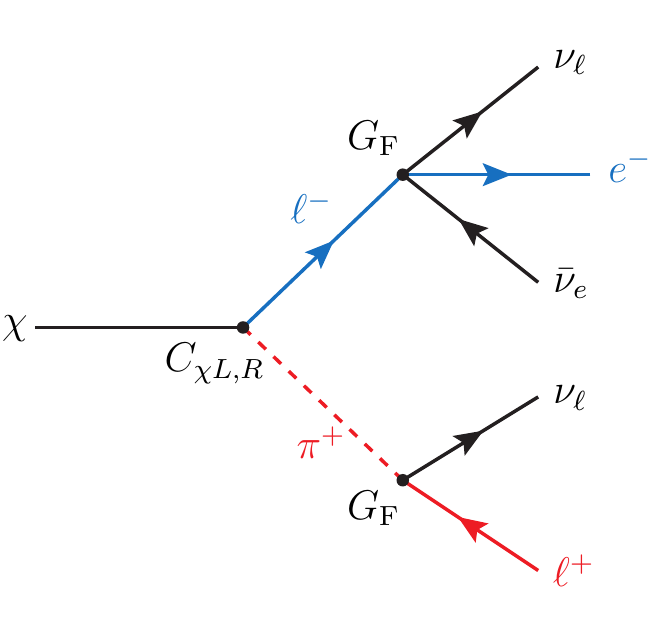}\qquad\qquad
\includegraphics[width=0.4\textwidth]{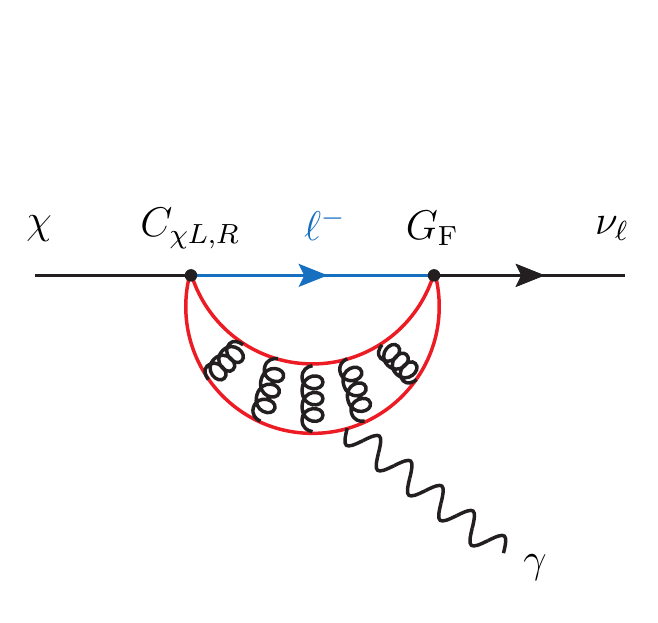}
\caption{Decays of DM coupled to $\mu\pi$. {\bfseries Left:} via off-shell $\mu$ and/or $\pi$.
{\bfseries Right:} at loop level. The $q\bar q$ quarks (in red) form QCD bound states, including pions.
\label{fig:Feyndecays}}
\end{figure}

The electron and proton are stable in view of conservation of charge and baryon numbers.
The other SM particles undergo weak decays suppressed by $(m/v)^4$ and possibly by phase space
(the neutron) or by flavour (the $K, \Omega, \Sigma$ in table~\ref{tab:narrow}).
DM decays can be suppressed by  $\epsilon \sim10^{-15}$, far from the needed $\epsilon < 10^{-30}$.
One dose of off-shell vaccine does not protect DM stability well enough.
Two doses of the off-shell vaccine can give enough protection.
Let us assume that DM is coupled to two SM particles
as in eq.\eq{DM12} and its mass is $M < m_1,m_2$.
Then doubly-virtual tree-level decays
$ {\rm DM} \to {\rm SM}_1^*\, {\rm SM}_2^*$
are suppressed by $\epsilon_1 \epsilon_2 \sim 10^{-30}$,
corresponding to two weak interactions.
An example in this sense is DM coupled to $\mu\pi$, see fig.\fig{Feyndecays}a.


\bigskip

However, even if DM only directly couples to massive SM particles, 
so that its tree-level decay channels are suppressed as discussed in the previous section,
one-loop effects can give DM decays into light SM particles.
DM that couples to a particle/anti-particle pair (such as $\pi^+\pi^-$) decays
at one loop level into photons, ${\rm DM}\to\gamma\gamma$, with rate
suppressed by a QED loop factor, $\epsilon \sim  (e/4\pi)^4 \sim 10^{-5}$.
This is far away from the needed $10^{-30}$ suppression.
DM that couples to $\pi ^\pm\ell^\mp$  decays as ${\rm DM}\to \nu \gamma $ at one loop level
(see fig.\fig{Feyndecays}b), with rate
suppressed by an electroweak loop factor, $\epsilon \sim (f_\pi/4\pi v)^4 \sim 10^{-18}$,
but still below the needed $10^{-30}$.
Loop decays are present because this DM has lepton number,
which is carried by nearly-massless neutrinos. 
In section~\ref{DM2} we discuss non-minimal 
models that achieve the level of stability needed for freeze-out via thermal
decays. We here discuss a minimal model 
that achieves the level of stability needed for freeze-in.

\subsection{Mediator models}\label{DMellpi}
We consider DM to be a fermion  $\chi$ with no SM gauge interactions and lepton number one.
This corresponds to a sterile neutrino in presence of a Yukawa coupling $\chi LH$.
This is a well-known allowed DM candidate if it has keV-scale mass:
it can be cosmologically produced,
compatibly with bounds on $\chi\to \nu \gamma$ decays~\cite{DMreview}, as freeze-in, via resonant oscillations or via the decay of an extra scalar.

We instead explore the possibility that the SM neutral lepton singlet $\chi$ has a heavier mass $M$ and has 
gauge interactions in an extension of the SM.
The natural possibilities allowed by group theory are the following:

\begin{itemize}
\item A $(\bar \chi \gamma_\mu \ell_R )\,W_R^\mu$ interaction in models where the right-handed $W_R$ boson
is part of an extra $\SU(2)_R$. The full couplings of the $W_R$ mediator to fermions are
\beq \label{eq:LQ}
\mathscr{L}_{W_R} = \left( 
g_\chi^i \, \bar \ell_R^i \gamma_\mu \chi  +
g_R^{ij}\, \bar u^i_R \gamma_\mu d^j_R \right) W_R^\mu  + \hbox{h.c.} \eeq
where $i,j$ are flavour indices. 
Integrating out the $W_R$ generates
\beq \label{eq:DMqqlWR}
\Lag^{\rm eff}_{W_R}   =  -C_{\chi R} (\bar\chi \gamma^\mu  \ell_R) (\bar d_R \gamma_\mu u_R) + \cdots
\eeq
with $C_{\chi R} = g_\chi g_R/M_{W_R}^2$.

\item A $(\bar u_R\gamma_\mu   \chi)\, {\cal U}^\mu$ interaction with a vector leptoquark $\mathcal{U} \sim ({ 3},{ 1})_{2/3}$~\cite{Fajfer:2015ycq,Buttazzo:2017ixm}, that can arise in models with a $\SU(4)$ Pati-Salam symmetry~\cite{Pati:1974yy,1512.01560,Barbieri:2017tuq,DiLuzio:2017vat}.
Its couplings to fermions are
\beq \label{eq:LQ}
\mathscr{L}_{{\cal U}} = \left( 
g_\chi^i \, \bar u_R^i \gamma^\mu \chi  +
g_R^{ij}\,  \bar d_R^i \gamma^\mu \ell_R^j +
g_L^{ij}\, \bar Q^i_L \gamma^\mu L^j_L \right) {\cal U}_\mu  + \hbox{h.c.} \eeq
We will discuss in 
section~\ref{flavour} how a flavour structure can arise.
The latter term has been considered as a possible interpretation of the 
flavour anomalies in decays of $B$ mesons,
adding interest to the possibility of obtaining DM in this context.
Integrating out the leptoquark at tree level generates
\beq \label{eq:LeffU}
\Lag^{\rm eff}_{\cal U}=  - C_{\chi R} (\bar\chi \gamma^\mu  \ell_R) (\bar d_R \gamma_\mu u_R) +
2C_{\chi L} (\bar\chi L_L)(\bar Q_L u_R)-
C_{LL} (\bar s_L \gamma^\mu b_L)(\bar\mu_L \gamma_\mu\mu_L)+\cdots\eeq
with $C_{\chi R}=\sfrac{g_\chi g_R}{M_{\cal U}^2}$,
$C_{\chi L}=\sfrac{g_\chi g_L}{M_{\cal U}^2}$, and
$C_{LL} = \sfrac{g_L^{22} g_L^{32}}{M_{\cal U}^2}$,
which reproduces the $B$-anomalies for 
$C_{LL} \approx  -1.4\, V_{tb} V_{ts}^* \alpha_{\rm em}/4\pi v^2 = 1.4 /(36\TeV)^2$~\cite{2103.13370}.
The second operator in eq.~\eqref{eq:LeffU} can also be generated by the exchange of a vector leptoquark $\tilde{\mathcal{V}}_2\sim (3,2)_{1/6}$, with coefficient $C_{\chi L} = g_\chi g_L/M_{\tilde{\mathcal{V}}_2}^2$ and couplings
\begin{equation}\label{V2}
\Lag_{\tilde{\mathcal{V}}_2} =  \left(g_L \bar u_R^c \gamma_\mu \varepsilon \ell_L + g_\chi \bar q_L^c \gamma_\mu \chi \right)\tilde{\mathcal{V}}_2^\mu + \text{h.c.}
\end{equation}
A vector leptoquark $\bar{\mathcal{U}}_1\sim (3,1)_{-1/3}$ can couple to $\chi$ and $d_R$, but not to other SM currents, and thus cannot generate any DM-SM$^3$ interaction.

\item Alternatively, interactions can be mediated by a variety of scalar fields. An example is the already mentioned Yukawa coupling with a Higgs doublet, $H^c \bar L_L \chi$, in which case the DM can be identified with a right-handed neutrino~\cite{DMreview}. Scalar leptoquarks with suitable quantum numbers, instead, can couple $\chi$ to SM quarks~\cite{Buttazzo:2017ixm,Bauer:2015knc,Gherardi:2020qhc,Dorsner:2016wpm}. The only two possibilities that are able to induce an interaction with SM leptons are $S_1 \sim (3, 1)_{-1/3}$, and $\tilde R_2 \sim (3, 2)_{1/6}$, with
\begin{align}
\Lag_{S_1} &= (g_\chi \bar d_R \chi + g_L \bar Q_L \varepsilon L_L^c + g_R \bar u_R e_R^c) S_1 + \text{h.c.},\label{S1}\\
\Lag_{\tilde R_2} &= (g_\chi \bar Q_L \chi + g_L \bar d_R L_L \varepsilon ) \tilde R_2 + \text{h.c.}\label{R2}
\end{align}
In eq.~\eqref{S1} and \eqref{V2} above we omitted di-quark couplings for $S_1$ and $\tilde{\mathcal{V}}_2$ that would violate baryon number. A scalar leptoquark $\bar S_1\sim (3,1)_{2/3}$ can also couple to $\chi$ and have di-quark couplings, thus inducing a mixing between $\chi$ and the neutron.
\end{itemize}
The full list of possible DM couplings to one lepton and two quarks, all of which can be generated by the tree-level exchange of one of the previous mediators,
 is~\cite{0806.0876}
\be
\begin{aligned}  \label{eq:DMqql}
\Lag^{\rm eff}_{\rm DM} =
&-C_{\chi R}(\bar \chi \gamma^\mu \ell_R)(\bar d_R\gamma_\mu u_R)
+2C_{\chi L} (\bar\chi L_L)(\bar Q_L u_R)\\
&-C_{\chi L}'(\bar L_L\chi)\varepsilon(\bar Q_L d_R)
-C_{\chi L}''(\bar Q_L\chi)\varepsilon(\bar L_L d_R). 
\end{aligned}
\ee
We expect that the various operator coefficients satisfy $C_i \circa{<} ({\rm TeV})^{-2}$ because 
the charged or colored mediators that generate the effective interactions are generically constrained to be heavier than the TeV scale by collider bounds. 
We will  sometimes collectively denote these couplings as $4\GDM/\sqrt{2}$,
including a normalization factor for ease of comparison with the Fermi constant $\GF$.

\smallskip

Below the QCD scale $ \Lag^{\rm eff}_{\rm DM}$  becomes effective couplings to pions, $\chi \ell \pi$, 
and to heavier mesons (such as $\rho$ and excited pions), as well as couplings to baryons,
$\chi \ell \bar p n$. Heavier mesons give negligible effects in cosmology and are not subject to 
significant bounds, despite that kinematical space for their tree-level decays is open.
The coupling to baryons can have more significant effects, since baryons remain as relics at $T \ll m_{p,n}$
thanks to the baryon asymmetry (DM couplings to quarks never lead to
DM coupled to baryons but not to pions).
Using chiral perturbation theory (see appendix~\ref{appA}) the coupling to pions is obtained substituting the quark terms in the
effective operator eq.\eq{DMqql} with pion terms with the same transformation properties:
\begin{eqnarray}
\Lag_{\rm DM}^{\pi}   &=&\frac{C_{\chi R}}{\sqrt{2}}  (\bar\chi\gamma^\mu \ell_R)\Big[f_\pi\, D_\mu\pi^+ + i (\pi^0\partial_\mu\pi^+ - \pi^+\partial_\mu\pi^0)+\nonumber
\cdots\Big]+ \\
&&- i C_{\chi L} f_\pi B_0 \Big[ \sqrt{2} (\bar\chi \ell_L) \pi^+ + (\bar\chi \nu_L) \pi^0 + \cdots\Big] + \text{h.c.}
\label{eq:LagDMellpi}
\end{eqnarray}
where  $f_\pi =93\MeV$, $B_0 \approx 22 f_\pi$ is the quark condensate,
$D $ is the gauge-covariant derivative and we will not need higher orders.
We assumed DM couplings to right-handed quarks, but this information is
lost in DM coupling to pions (the suppression is of order $f_\pi/\Lambda_{\rm QCD}\sim 1$).
This will remove a suppression of weak loops.

\subsection{DM decays}\label{ellpidecays}
We start discussing loop decays into massless particles, which are always kinematically allowed, and then move to tree-level decays.

The electroweak interactions couple pions to the charged lepton current (the SM Lagrangian is written in eq.\eq{SMpi}).
As a result leptonic DM with the interactions of eq.\eq{DMqql} acquires at one loop 
a negligible mass mixing between DM and neutrinos\footnote{In the pion effective theory 
the mass mixing $ \mu \, \chi \nu + \hbox{h.c.}$ is UV-divergent and estimated as
\beq\qquad\mu \sim \frac{C_{\chi R}    f_\pi^2 m_\ell g_2^2 \Lambda_{\rm UV}^2}{(4\pi M_W)^2} \sim {\rm meV} \frac{m_\ell}{m_\mu} \frac{C_{\chi R} }{\GF} . \eeq
The UV cut-off has been estimated as $\Lambda_{\rm UV}^2 \sim M_W^2/(4\pi)^2$.
Indeed, in the high-energy theory above the QCD and EW scales, 
the analogous effect is a log RG mixing between two dimension 6 operators:
 eq.\eq{DMqql} at two loop induces $\chi LH H^\dagger H$, so
 $\mu \sim C_{\chi R}  v^3 \sfrac{y_\ell y_u y_d g_2^2}{(4\pi)^4}$. 
 The extra $H H^\dagger$ and thereby $y_u y_d$ terms appear because, in the full theory, DM is coupled to right-handed quarks. },
and a mixed magnetic moment with neutrinos:\footnote{Above the QCD and EW scales, the analogous effect is negligible: the dimension-6 operator of
eq.\eq{DMqql} induces at 2 loops the operator $(\bar\chi \sigma_{\mu\nu} LH)B^{\mu\nu} HH^\dagger$.}
\beq \Lag_{\rm eff}=  Ae( \bar \chi \sigma_{\mu\nu} \nu)F_{\mu\nu}\qquad\hbox{with}\qquad
A \approx    \frac{\GF   \Lambda_{\rm QCD}^2}{(4\pi)^2 } (m_\ell  C_{\chi R} + \Lambda_{\rm QCD} C_{\chi L})
.  \eeq
A precise estimate is not possible, as the exact loop integral is dominated by momenta
around the QCD scale where all resonances contribute.
The resulting DM decay rates are
\beq \label{eq:DMnugammaR} 
\Gamma(\DM \to \nu\gamma) = \frac{e^2 A^2 M^3}{2\pi} \sim \frac{1}{4\,{\rm yr}}\frac{M^3}{m_\pi^3} 
\left( \frac{C_{\chi L}  + C_{\chi R}  m_\ell / \Lambda_{\rm QCD}}{\GF }\right)^2.
\eeq

\bigskip

  \begin{figure}[t]
\begin{center}
\includegraphics[width=\textwidth]{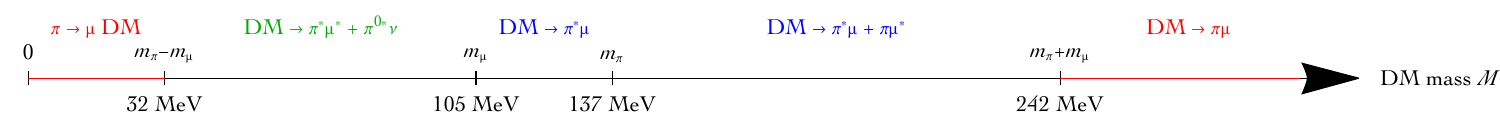} \end{center}
\caption{\label{fig:kinematics} DM with lepton number coupled to $\mu\pi$:
kinematical thresholds for most dangerous decays.
A $*$ denotes a virtual particle, and thereby a $\Gamma/M$ suppression.
We here neglected the mass difference between charged and neutral pions.}
\end{figure}

\begin{figure}[t]
\begin{center}
\includegraphics[width=0.44\textwidth]{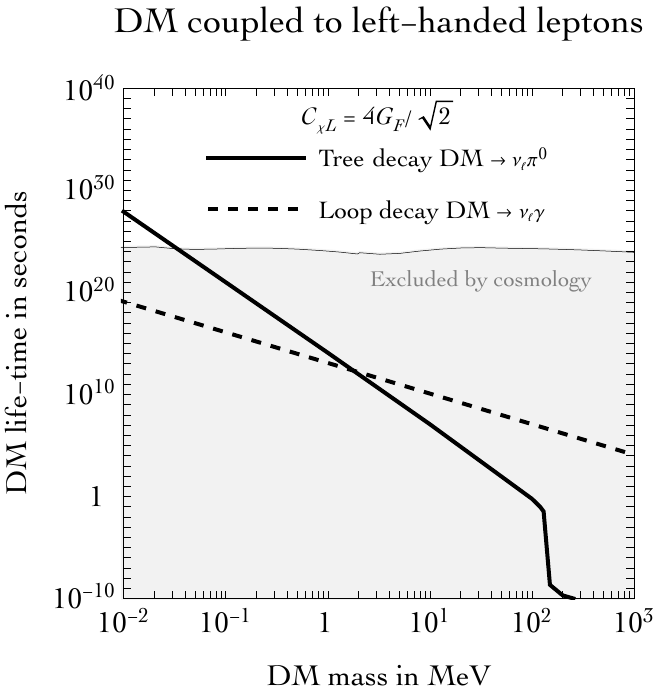} \qquad
\includegraphics[width=0.44\textwidth]{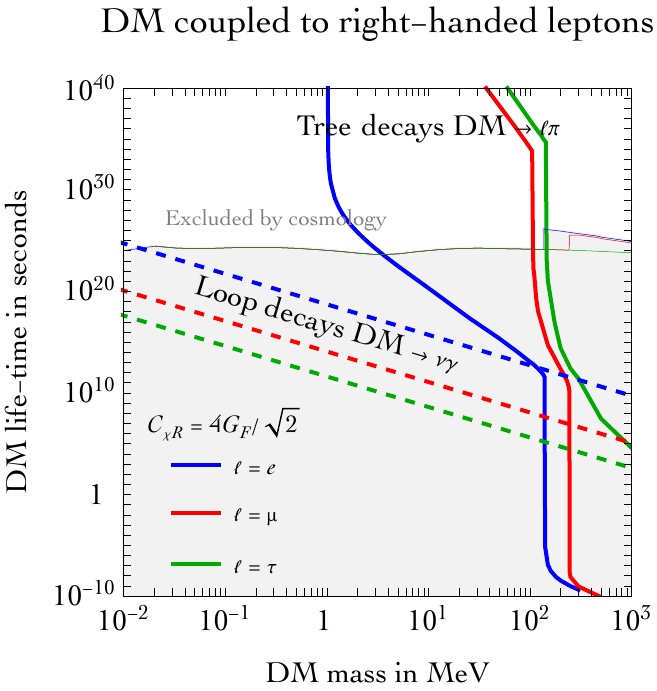}
\end{center}
\caption{\label{fig:ViteMedieDM} Life time of DM $\chi$ coupled to a 
lepton $\ell$ and to quarks.
{\bfseries Left}: DM coupled to left-handed leptons
with coupling $C_{\chi L}=4\GF/\sqrt{2}$.
The dashed curve is the contribution from DM loop decays into $\nu_\ell \gamma$; 
the continuous curve is the contribution from tree-level decays into $\nu_\ell \pi^0 \to \nu_\ell\gamma\gamma$.
{\bfseries Right:} DM coupled to right-handed leptons with coupling $C_{\chi R}=4\GF/\sqrt{2}$.
Curves can be rescaled for different couplings taking into account that life-times scale as $1/C_{\chi L,\chi R}^2$,
while the bounds from cosmological DM stability and from CMB do not depend on DM couplings,
and are discussed in the text.}
\end{figure}

Coming to tree-level decays,
they can a) dominate over loop decays if $\DM\to \ell \pi$ is kinematically open, or
b) be comparable to loop decays if one of the two SM particles needs to be off-shell to kinematically open the decay, or
c) be negligible compared to loop decays if both $\ell^*\pi^*$ need to be off-shell.
Let us discuss in more detail the case of DM coupled to $\ell=\mu$ as a concrete example. 
Fig.\fig{kinematics} shows the main kinematical regimes.
In the co-stability range 
\beq 34\MeV = m_\pi - m_\mu <M<m_\pi + m_\mu =245\MeV\eeq
all tree-level decays among $\chi,\pi,\mu$ are closed.
DM heavier than $m_\pi+m_\mu$ decays at tree level as $\DM\to\mu^\pm\pi^\mp$.
One of these particles must be off shell if the DM mass $M$ is in the range $m_\mu < M < m_\pi+m_\mu$,
suppressing the DM tree-level decay rate by a $\epsilon_\mu,\epsilon_\pi \sim 10^{-17}$ factor.
Furthermore, $\pi\to \mu^* \chi \to e\bar\nu_e \nu_\mu \chi$ 
decays are suppressed by $\epsilon_\mu\sim 10^{-16}$,
and $\mu\to \pi^* \chi = e \bar\nu_e \chi$ decays by $\epsilon_\pi \sim 10^{-19}$.
So bounds $\hbox{BR}(\pi \to \mu \chi)\circa{<}10^{-7}$
are easily satisfied.
If $m_\pi - m_\mu < M < m_\mu$ (in green in fig.\fig{kinematics}), the tree-level $\DM \to \mu^*\pi^*$ decay rate 
is suppressed by $\epsilon_\pi \epsilon_\mu \sim 10^{-33}$ as
both $\mu$ and $\pi$ must be off-shell (see appendix~\ref{DMellpitreedecay} for details).
However, in this case, loop decays dominate.

\medskip
Fig.\fig{ViteMedieDM} shows the life-time of DM coupled to left-handed (left panel) and right-handed (right panel) leptons  $\ell$ and  quarks. 
We fix a reference value for $C_{\chi R}$ and $ C_{\chi L}$ equal to $4\GF/\sqrt{2}$. 

\smallskip 

The solid curve in the left panel is the contribution from tree-level decays $\DM\to \nu_\ell \pi^0$. 
For $M<m_{\pi_0}$ this channel is kinematically closed and therefore the relevant contribution is  into $  \nu_\ell\gamma\gamma$ via the off-shell $\pi_0$. 
The dashed curve is the  contribution from DM loop decays $\DM \to \nu_\ell \gamma$ which is independent 
on the lepton mass, as shown in eq.~(\ref{eq:DMnugammaR}), 
and is the dominant decay channel for $M<2$ MeV.

In the right-panel of the same figure, the solid curves indicate  the contributions from tree-level decays, $\DM\to \ell \pi$. 
These predictions depend on the leptons (blue, red and green are for $e$, $\mu$ and $\tau$ respectively) 
and the different features are due to the kinematical thresholds as discussed in details above. 
The dashed curves show the  contribution from DM loop decays into $\nu_\ell \gamma$, which depend 
on the lepton mass since we need a mass insertion to flip the chirality of the lepton (see eq.~\ref{eq:DMnugammaR}). 
For a given lepton the loop decay is dominant at $M<m_\ell + m_\pi$ and therefore the co-stability 
regime is not important from the phenomenological point of view. 

\smallskip

For life-times longer then the age of the Universe, limits from {\sc Planck} and {\sc Voyager II} apply.
These bounds are shown as grey areas in the figure. For DM coupled to left-handed leptons, since there is always a 
neutrino that does not release energy into the plasma, we rescale the {\sc Planck} bound for the channel DM $\to \gamma\gamma$ 
given in fig.~7 of~\cite{1610.06933}. 
In particular for $M>2$ MeV we rescale the limit by a factor 2/3 (the final state is $\nu_\ell \gamma\gamma$ for both on- and off-shell $\pi_0$), while for lighter 
DM by a factor 1/2 since the loop decay into $\nu_\ell \gamma$ dominates. 
The bound is more complicated for DM coupled to right-handed leptons.  
For $M<m_\ell + m_\pi$ the dominant channel is
into $\nu_\ell \gamma$, so we follow the same procedure described above. 
For  $M>m_\ell + m_\pi$ the final 
state after the decay of the pion and lepton always involves a $e^-e^+$ pair with a given number of neutrinos (3, 5 and 7 for $e$, $\mu$ and $\tau$ respectively). 
We use the {\sc Planck} and {\sc Voyager II}  limits given in~\cite{1612.07698,2008.01084} for the decay channel 
$\DM \to e^+ e^-$ rescaling them by a factor 2/5, 2/7 and 2/9 for $e$, $\mu$ and $\tau$ respectively. 
This is a good approximation for {\sc Planck}, since the deposited energy matters, while
for  {\sc Voyager II}  this is a rough approximation given that the energy spectrum of the $e^\pm$
changes dramatically as we have a chain of multiple decays.

\begin{figure}
$$\begin{array}{cc}
\hbox{process} & \hbox{Boltzmann factor }\Delta\\ \hline \rowcolor[rgb]{0.95,0.97,0.97}
\chi \ell^\pm \leftrightarrow \pi^\pm \gamma & \max(m_\ell,m_{\pi^\pm}-M) \\  \rowcolor[rgb]{0.95,0.97,0.97}
\chi \gamma \leftrightarrow\ell^\pm \pi^\mp & m_\ell + m_{\pi^\pm}-M\\ \rowcolor[rgb]{0.95,0.97,0.97}
\chi  \pi^\pm \leftrightarrow\ell^\pm   \gamma &m_{\pi^\pm} \\ 
\hline  \rowcolor[rgb]{0.97,0.97,0.93}
\chi \pi^\pm \leftrightarrow\ell^\pm \pi^0 & \max(m_{\pi^\pm}, m_\ell + m_{\pi^0} -M)   \\ \rowcolor[rgb]{0.97,0.97,0.93}
\chi \pi^0 \leftrightarrow\ell^\pm \pi^\mp &  \max(m_{\pi^0}, m_\ell + m_{\pi^\pm} -M)  \\ \rowcolor[rgb]{0.97,0.97,0.93}
\chi \ell^\pm \leftrightarrow \pi^\pm \pi^0 & \max(m_\ell , m_{\pi^0}+m_{\pi^\pm}  -M)  \\ 
 \hline \rowcolor[rgb]{0.99,0.97,0.97}
\bar\chi \chi \leftrightarrow \ell^-\ell^+ & \max(M ,2 m_\ell - M) \\ \rowcolor[rgb]{0.99,0.97,0.97}
\bar\chi \chi \leftrightarrow \pi^-\pi^+ & \max(M ,2 m_\pi - M)
\end{array}\qquad
\raisebox{-15ex}{\includegraphics[width=0.49\textwidth]{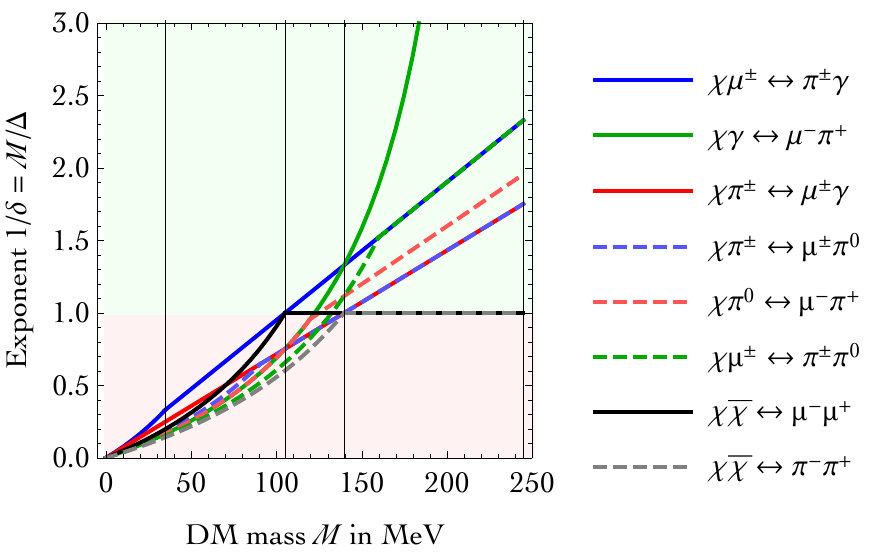}}$$
\caption{\label{fig:Deltaexponents}Assuming DM coupled to $\ell\pi$, we list the
main processes that change by one unit the DM number.
Their contribution to cosmological DM freeze-out is controlled by the exponent
$\delta = \Delta /M$, plotted in the figure.
Standard DM annihilations correspond to exponent $\delta =1$, 
and more efficient scatterings correspond to $\delta< 1$.}
\end{figure}

\subsection{DM thermal decays}\label{ellpi}
We can neglect scatterings that involve two DM particles, as their cross sections $\sigma_2$ arise at second order in the
small new physics couplings $C$ of eq.\eq{LagDMellpi}.
On the other hand, $2\leftrightarrow 2$ scatterings that change DM number by one unit arise at first order, by
combining the $C \,\chi \ell \pi$ interaction with either 
\begin{itemize}
\item an electro-magnetic interaction with coupling $e \sim 0.3$, 
adding an extra $\gamma$ to the process (rates in Appendix~\ref{extragamma}); or
 
\item a strong interaction with coupling $\sim m_\pi/f_\pi \sim 1$ 
(see eq.\eq{LagDMellpi}; the SM pion Lagrangian contains no $\pi^3$ interactions),
adding an extra massive $\pi$ to the process. The rates are given in Appendix~\ref{extrapi}.

\end{itemize}
So these processes have a larger cross section $\sigma_1 \gg \sigma_2$, that needs a much smaller critical value
$\sigma_1 > \sigma_1^{\rm cr}$
to affect DM cosmological freeze-out, if their kinematical factors $\Delta$ 
that control the Boltzmann suppression are smaller than the DM mass $M$.
The most important processes are shown in fig.\fig{Deltaexponents}.
It shows that some off-shell decays can be less Boltzmann suppressed than DM annihilations,
but only when $M >m_\ell$, so that tree-level DM decays are not suppressed by two off-shell factors.\footnote{A similar situation is found if DM with lepton number 1 couples instead to $\ell =e$: 
less Boltzmann suppressions needs $M > m_\pi/2$, so that 
$\DM \to \pi^* e$ 
are kinematically open and too fast.}

\begin{figure}[t]
\begin{center}
\includegraphics[width=0.4\textwidth]{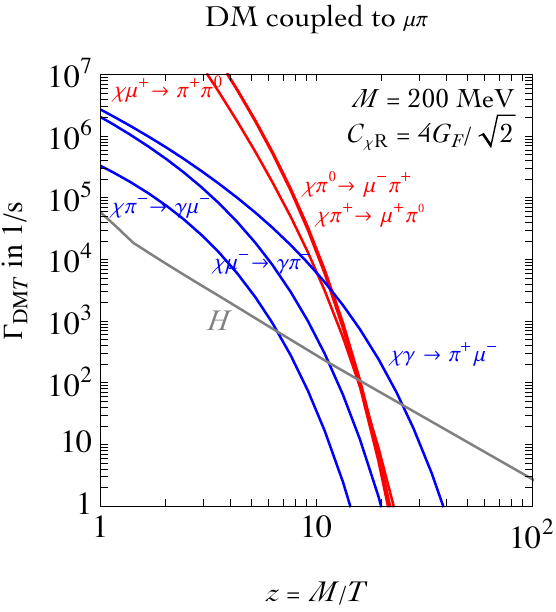} \qquad
\includegraphics[width=0.42\textwidth]{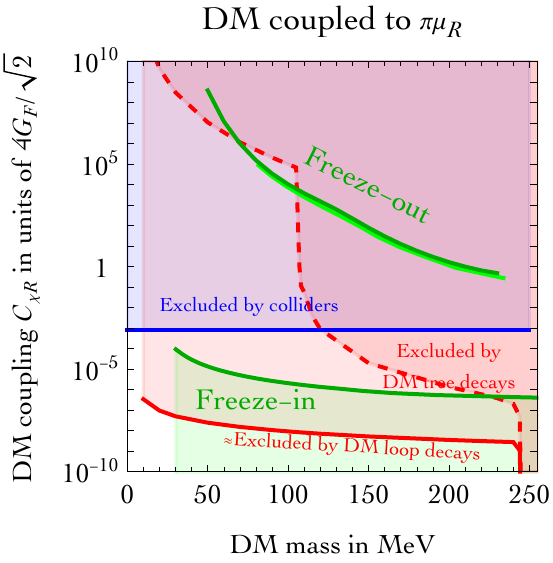} \end{center}
\caption{Dark Matter as a fermion coupled to $\mu^\pm\pi^\mp$.
{\bfseries Left}: rates of main thermal decays compared to the Hubble rate.
As expected, $\chi \gamma\leftrightarrow \pi\mu$ dominates at $T \ll M$.
{\bfseries Right}: parameter space (mass, coupling).
The red region is excluded by DM decays at tree (dashed) or loop (continuous) level,
estimated as in eq.\eq{DMnugammaR}.
The blue region is excluded by collider data constraining the quark-level effective operators of eq.~\eqref{eq:DMqql} as explained in section~\ref{sec:collider}.
The DM abundance is reproduced thermally along the upper green curve,
or as freeze-in below the lower green curve. 
\label{fig:DMmupiSummary}}
\end{figure}

Fig.\fig{DMmupiSummary} shows the final result.
Its left panel is reported the thermal decay rates as function of $z=M/T$:
as expected the process with smallest $\delta$  
($\chi\gamma\leftrightarrow \pi \mu$ in the example) dominates at low temperature.
The green curves in the right panel show the coupling needed to reproduce the cosmological DM abundance 
as freeze-out via thermal decays (one curve is the numerical result, one nearby curve is the analytic approximation).
It shows that DM interactions to $\mu\pi$ needs to be a few order of magnitude stronger than weak interactions.
As expected, freeze-out is excluded by too fast DM decays~\cite{1612.07698,2008.01084} (red curve).
%

\subsection{Collider bounds}
\label{sec:collider}
Let us discuss collider bounds on the effective $(\bar \chi \gamma_\mu q)(\bar q \gamma^\mu \ell)$ interaction.
We find that the strongest constraints are set by the ATLAS analysis in \cite{1906.05609}, which perform a search for a heavy charged boson in events with a charged lepton and missing transverse momentum.
In order to derive the bound, we have performed a recast of the ATLAS analysis. In particular, we have considered the reported bounds from the model-independent analysis in the muon channel, based on cuts on the variable $m_T$, defined as
$
m_T = [2 p^\ell_T E^{\rm miss}_T ( 1 - \cos\phi(\ell, E^{\rm miss}_T))]^{1/2}$.
In order to identify the fiducial region, we selected:
\begin{equation}
p^\ell_T> 55 \, \text{GeV}\; ,\qquad  |\eta|_\ell < 2.5 \; ,  \qquad E^{\rm miss}_T> 55 \, \text{GeV}\; ,\qquad  m_T> 110 \, \text{GeV} \, .
\end{equation}
We considered the model where a leptoquark, with mass $M_{\rm med} \gg \sqrt{s}$, mediates the effective interaction in eq.~(\ref{eq:DMqql}), obtaining the constraint:
\begin{equation}\label{eq:CchiRcoll}
C_{\chi R}< \frac{4\GF}{\sqrt{2}} \left\{\begin{array}{ll}
8.1 \times 10^{-4}\; & \text{for } \ell=\mu,\cr
 5.9 \times 10^{-4} \; &\text{for } \ell= e.\end{array}\right.
\end{equation}
These bounds surpass the limits from HERA~\cite{Cornet:1997vy}.  Calculations have been made with {\sc MadGraph}~\cite{1405.0301}. Note that the limits are obtained in the effective field theory approximation  and we thus expect $\mathcal{O}(1)$ corrections in the lower mass region where $M_{\rm med} \approx \sqrt{\hat{t}} \approx$~TeV, with $\sqrt{\hat{t}}$ denoting the energy exchanged in the $t$-channel propagator of the leptoquark. This result also excludes thermal freeze-out, and is plotted as a blue curve in fig.\fig{DMmupiSummary}.

\smallskip

This conclusion still holds taking into account DM scatterings with baryons (such as $\chi p \to n \ell^+$),
that do not get Boltzmann suppressed at low $T$ thanks to the baryon asymmetry $Y_B\approx 0.8\cdot 10^{-10} $.
Similarly to weak interactions of neutrinos, these processes maintain DM in thermal equilibrium down to
$T \sim (M_{\rm Pl} Y_B \GDM^2)^{-1/3} $.
However, if $C \circa{<} \GF $, 
this decoupling temperature cannot be below the desired DM mass $M$.

\subsection{DM relic abundance via freeze-in}
While freeze-out is excluded, freeze-in is allowed.
Three different regimes are possible, depending on the highest temperature $T_{\rm RH}$ attained by the SM plasma:
\begin{enumerate}
\item At $T \circa{<}\Lambda_{\rm QCD} $
in the confined phase, DM interacts with pions according to eq.\eq{LagDMellpi}. The dimension-5 interaction with right-handed leptons
contributes to the DM freeze-in abundance
as $Y \sim e^2 C_{\chi R} ^2 f_\pi^2 M_{\rm Pl} T$:
the DM abundance is dominated by the maximal temperature $T_{\rm RH}$.
The minimal amount of freeze-in (upper boundary of the green region in fig.\fig{DMmupiSummary}b)
arises if $T_{\rm RH}\approx M$.
This confined contribution alone allows to reproduce the DM abundance for
$C_{\chi R} \sim \sqrt{T_0/MM_{\rm Pl}}/ef_\pi^{3/2} \sim 10^{-6}\GF $,
nearly compatible with DM stability for $M < m_\pi+m_\mu$. For the renormalizable interaction with left-handed leptons, freeze-in happens around $T\sim M$, and the abundance $Y\sim e^2 C_{\chi L}^2 f_\pi^2 B_0^2 M_{\rm Pl}/M$ is independent of $T_{\rm RH}$.
The pion description breaks down at $T \sim \Lambda_{\rm QCD}$ when the QCD phase transition happens.

\item
At temperatures above the QCD phase transition and below the mass $M_{\rm med}$
of the particle that mediates the dimension-6  effective  interaction with quarks in eq.\eq{DMqql}, 
it leads to a higher $Y \sim  C_{\chi R} ^2 M_{\rm Pl} T^3 $.
This is again dominated by the highest temperature $T \circa{<}\min (T_{\rm RH}, M_{\rm med})$,
where $M_{\rm med} \circa{<}C_{\chi R} ^{-1/2}$.
Since very small DM couplings are needed in order to get the cosmological DM abundance, 
we can neglect annihilation processes that involve two DM particles,
described by extra effective operators of the type $|\bar u_R \gamma_\mu \chi|^2$.

\item At temperatures above the mediator mass, DM has renormalizable interactions
and freeze-in is dominated at $T \sim M_{\rm med}$.
The dimension-less DM coupling to the mediator must be small, in order 
not to over-produce DM.
Other interactions (such as gauge interactions with SM particles) can be large and keep the mediator in thermal equilibrium,
so that the freeze-in DM abundance is produced by its decays.
In Boltzmann approximation, the space-time density of mediator decays into DM is 
\beq
\gamma_{\rm eq} = n_{\rm med}^{\rm eq} \frac{{\rm K}_1 (M_{\rm med}/T)}{{\rm K}_2 (M_{\rm med}/T)}\Gamma_{\rm dec}=
\frac{d_{\rm med}}{2\pi^2} \Gamma_{\rm med} T M^2_{\rm med} \,{\rm K}_1 (M_{\rm med}/T),\eeq
where $\Gamma_{\rm med}$ is the mediator partial decay width into DM,
and $d_{\rm med}$ is its number of degrees of freedom.\footnote{We can neglect
scattering rates because suppressed by extra powers of couplings, such as $g_2$.
At $T\gg M_{\rm med}$ the decay rate is suppressed by a Lorentz factor
$T/M_{\rm med}$ compared to the scattering rates $\gamma_{\rm eq} \sim g_2^2 g_\chi^2 T^4$.
This enhancement of scattering rates is not important for
the final DM abundance, because it is dominated at $T \sim M_{\rm med}$.
Furthermore this enhancement can be partially included as a NLO contribution to the decay rate, 
describing it as the thermal contribution to the mediator mass, $\delta M_{\rm med}^2 \sim g_2^2 T^2$.
The thermal mass avoid an apparent IR-enhancement at $T \gg M_{\rm med}$
of $t$-channel scattering rates, that would anyhow cancel with loop corrections in a full NLO computation
(KNL theorem, see~\cite{1106.2814,2107.07132} for a similar computation).}
Inserting $\gamma_{\rm eq}$ in eq.\eq{FIestimate} gives
\beq Y_\infty = \frac{405\sqrt{5} M_{\rm Pl} d_{\rm med} \Gamma_{\rm med}}{16\pi^{9/2} d_{\rm SM}^{3/2} M_{\rm med}^2}.\eeq
\end{enumerate}
If the mediator is the ${\cal U}$ leptoquark 
one has $\Gamma_{\rm med} = \sfrac{g_\chi^2 M_{\rm med}}{24\pi }$ and $d_{\rm med}=18$,
and the DM abundance is matched for
\beq \label{eq:gchiexpr} g_\chi \approx 2.2~10^{-11}\sqrt{\frac{\GeV}{M}\frac{M_{\rm med}}{\TeV}} \,.
\eeq 
We remind that the coefficients $C$ of the effective Lagrangian depend also on other couplings of the mediator.
We now explore the possible flavour structure of these couplings, to see if some natural explanation can be found for the small size of $g_\chi$.

\section{Signals and anomalies}\label{signals}
\subsection{Flavour expectations for DM couplings and the $B$ anomalies}\label{flavour}
The fact that a leptoquark such as $\mathcal{U}_\mu$ could mediate at the same time the semi-leptonic interactions responsible for the $B$-physics anomalies, and the interactions responsible for DM freeze-in, although with couplings of different sizes, makes it interesting to speculate about a common origin of the two processes.

The $B$-physics anomalies can be explained, compatibly with other bounds from flavour and collider physics, in models where new physics couples dominantly to the third generation of fermions~\cite{1512.01560,Buttazzo:2017ixm,1609.09078,1506.01705,1806.07403,1506.02661}. Couplings to lighter generations of quarks and leptons then arise from the flavour rotations that diagonalize the Yukawa interactions, and are suppressed similarly to the masses and mixings of light fermions. 
This can applied to vector leptoquarks assuming that only 3rd-generation fermions
are charged under an extended Pati-Salam group~\cite{1712.01368,Greljo:2018tuh,Fuentes-Martin:2020pww}.
In the left-handed quark sector these rotations are of the order of the CKM matrix. 

Given the tiny values of the coupling $g_\chi$ needed to reproduce the cosmological DM abundance via freeze-in, for DM masses in the MeV range, it is then natural to consider the DM $\chi$ as a right-handed neutral lepton of the first family.
Other generations of $\chi$, with larger couplings to SM, can be so heavy that they decay back fast to SM particles, without contributing to DM.

%
%

We parametrize the SM Yukawa couplings as
\begin{align}\label{yukawa}
Y_u^{ij} &\approx \epsilon_{Q_i} \epsilon_{u_j}, & Y_d^{ij} &\approx \epsilon_{Q_i} \epsilon_{d_j}, & Y_e^{ij} &\approx \epsilon_{L_i} \epsilon_{e_j},
\end{align}
where $\epsilon_{f_i}$ are small parameters carrying the flavour suppression of the couplings to the $i$-th family of $f = \{Q, u, d, L, e \}$, and the relations hold up to $\mathcal{O}(1)$ factors.
Here and in subsequent estimates we also ignore an overall coupling,
assumed to be of order one.
Such a pattern of Yukawa couplings is easily obtained in Froggatt-Nielsen scenarios~\cite{Froggatt:1978nt}, or in models with partial composite fermions~\cite{Kaplan:1991dc,Contino:2006nn,Agashe:2004cp}.
More stringent relations between the $\epsilon_{f}^i$ can be obtained in models with a larger flavour symmetry, such as Minimal Flavour Violation~\cite{Chivukula:1987py,Hall:1990ac,DAmbrosio:2002vsn}
or $\U(2)$ models~\cite{Barbieri:2011ci,Barbieri:2012uh}.

The values of $\epsilon_{Q,u,d}$ can be estimated from the CKM mixings and the quark masses. Indeed,
\be
V_{\rm CKM}^{ij} \approx \epsilon_{Q_i} / \epsilon_{Q_j}, \qquad\quad \epsilon_{Q_i} \epsilon_{u_i} \approx (y_u, y_c, y_t), \qquad\quad \epsilon_{Q_i} \epsilon_{d_i} \approx (y_d, y_s, y_b),
\ee
from which one gets
\be\label{epsilonq}
\epsilon_Q \sim (\lambda^3, \lambda^2, 1)\, \eta_q, \qquad\quad \epsilon_u \sim (\lambda^4, \lambda, 1)\, \eta_q^{-1}, \qquad\quad \epsilon_d \sim (\lambda^4, \lambda^3, \lambda^2)\,\eta_q^{-1},
\ee
where $\lambda \sim 0.2$ is the Cabibbo angle, and $\eta_q$ is an overall $\mathcal{O}(1)$ parameter. The parameters in the lepton sector can not be estimated as easily at this level, 
since the neutrino mixings are not necessarily directly related to the Yukawa couplings.

We can now make the assumption that the flavour structure of the new physics couplings is determined by the same flavour parameters that control the SM Yukawas. Let us focus on the case of the vector leptoquark $\mathcal{U} \sim ({ 3}, { 1})_{2/3}$ for definiteness.
The couplings of $\mathcal{U}$ to fermions are defined in eq.~\eqref{eq:LQ}. In terms of the flavour spurions $\epsilon$ those couplings read
\begin{align}\label{LQ_couplings}
g_L^{ij} &\approx \epsilon_{Q_i} \epsilon_{L_j}, & g_R^{ij} &\approx \epsilon_{d_i} \epsilon_{e_j}, & g_\chi^i &\approx \epsilon_{u_i} \epsilon_\chi,
\end{align}
where we have introduced an additional (small) parameter $\epsilon_\chi$ that controls the DM couplings.

The exchange of a vector leptoquark contributes to semi-leptonic left-handed interactions via the couplings $g_L$. These couplings can thus be partially determined by fitting the $b\to s\ell\ell$ anomalies~\cite{Buttazzo:2017ixm}. The contribution to the Wilson coefficient of the semi-leptonic operator $(\bar b_L \gamma^\mu s_L)(\bar \mu_L \gamma_\mu \mu_L)$
 reads
\be
\Delta C_9^\mu = -\Delta C_{10}^{\mu} = \frac{g_L^{32}g_L^{22}}{M_\UL^2} \approx V_{tb}^* V_{ts}
\frac{\eta_q^2 \epsilon_{L_2} \epsilon_{L_2}}{M_{\cal U}^2} ,
\ee
which fits the observed deviations if $(\epsilon_{L_2})^2 \eta_Q^2 / M_\UL^2 \approx (6\,{\rm TeV})^{-2}$. 
For $M_\UL \approx\TeV$ the muon couplings can be $\epsilon_{L_2} \sim \mathcal{O}(0.1)$.
The first-generation coupling is required to be small, $\epsilon_{L_1} \ll \epsilon_{L_2}$, in order to suppress new physics effects in $b\to s e^+ e^-$.
The right-handed couplings $\epsilon_e$ are then related to the charged lepton Yukawa couplings, $\epsilon_{L_i} \epsilon_{e_j} \sim (y_e,y_\mu,y_\tau)$, and one has
\begin{align}
\epsilon_{e_1} &\gtrsim 10^{-4}, & \epsilon_{e_2} &\sim \text{few} \times 10^{-3}, & \epsilon_{e_3} &\sim 10^{-2},
\end{align}
where we assumed $\epsilon_{L_1} \lesssim \mathcal{O}(10^{-2})$  (an order of magnitude smaller than the muon coupling) and $\epsilon_{L_3} \sim 1$.
%

Notice that all the right-handed mixings $\epsilon_{u,d,e}$ estimated above are smaller than the left-handed ones $\epsilon_{Q,L}$ (except for the case of the electron where no hint on the size of $\epsilon_{L_1}$ exists). In particular, this is consistent with the observation of sizable LQ-induced effects in left-handed currents, but the absence of large right-handed currents.

\smallskip

Finally, reproducing the cosmological DM abundance via freeze-in as in eq.~\eqref{eq:gchiexpr},
fixes the LQ coupling to DM, and in turn $\epsilon_\chi \sim g_\chi/\epsilon_u$. Two limiting scenarios can be envisaged,
depending on which up-quark flavour couples dominantly to DM.
If the dominant decay is with the top quark, given the hierarchy of \eqref{epsilonq},
for $M \gtrsim 10\,{\rm keV}$ and $M_\mathcal{U}\sim {\rm TeV}$
one needs $\epsilon_\chi \lesssim 10^{-8}$,
much smaller than the corresponding electron spurion.
If instead $\chi$ couples dominantly to the first family of quarks, due to the additional $\epsilon_{u_1}$ suppression one gets $\epsilon_\chi \sim 10^{-5}$,
which is comparable to the analogous factor for electrons. The latter scenario could be motivated if $\chi$ is identified with the first-family member of a flavour triplet of right-handed neutrinos $\chi^i$: in this case its dominant coupling can be to other first-family fermions if flavour violation is suppressed in the right-handed sector, as is the case for instance in MFV or $U(2)$ models.
%
%

\medskip

We finally discuss why the Higgs doublet $H$, unlike the leptoquark $\UL$,
cannot generate DM via freeze-in. 
The Higgs doublet too can couple to DM as
\be\label{eq:Higgs}
y_\chi^i \, H^* \bar L_L^i \chi
\ee
where we estimate the Yukawa coupling as
$y_\chi^i \approx \epsilon_\chi \epsilon_{L,i}$.
Thus Higgs decays $h\to \nu_L\chi$ can in principle contribute to DM freeze-in production.
However DM dominantly produced in this way is excluded, because
eq.\eq{Higgs} implies a too large mixing of DM with neutrinos,
$\theta_{i} \approx y^i_\chi v/M_\chi$.
This is problematic because it gives a contribution to neutrino masses and, more importantly, because it contributes to DM decays as
\beq \Gamma(\DM \to \nu_i \gamma) = \frac{9\alpha\GF^2 M^5}{256\pi^4}\theta_i^2
\approx\frac{1}{10^{25}\sec} \left(\frac{y_\chi^i}{10^{-10}}\right)^2 \left(\frac{M}{\keV}\right)^3.
\label{Hfreezein}
\eeq
Higgs freeze-in instead needs a Yukawa coupling $y_\chi \approx 7\times 10^{-8}\sqrt{{\rm keV}/M}$, which is slightly larger for masses $M\gtrsim$ few eV.
And ad-hoc scalar coupled to two DM particles was introduced in~\cite{hep-ph/0609081} to achieve freeze-in:
we have shown that this role can be played by the vector leptoquark motivated by flavour anomalies,
as its ${\cal U} u \chi$ couplings do not directly induce DM decay.

In any case, we need to assume that $y_\chi$ is small enough so that eq.~\eqref{Hfreezein} is satisfied and the Higgs plays no role in freeze-in. In this case, the corrections to neutrino masses do not exceed the experimental values and do not impose further constraints.
We point out that the required Yukawa coupling is smaller than the naive expectation $y_\chi \approx \epsilon_\chi \epsilon_L$ even in the case where flavour violation is suppressed in the lepton sector and the dominant coupling is to electrons. 
The need for small mixings between left- and right-handed neutrinos is a common issue in models where the leptoquarks arise from partial unification at the TeV scale, and most likely requires a more complex neutrino sector~\cite{Greljo:2018tuh,Fuentes-Martin:2020pww}.

\subsection{Direct detection signals}\label{DD} 
Direct detection experiments are sensitive to DM scattering with light degrees of freedom such as light quarks, gluons and electrons. 
We consider DM that interacts with SM leptons and quarks as 
described by eq.~\eqref{eq:DMqql}, and in particular on the first two 4-fermion interactions.
After matching the interaction Lagrangian to the nucleon level, 
DM charged- and neutral-current interactions with nucleons arise. 
We now briefly discuss the novel signatures in direct and indirect detection searches of DM induced by these interactions.
We anticipate that, although these signatures are potentially interesting, the effective scale of the dimension 6 operators 
in eq.\eq{DMqql} must be large to avoid bounds from DM decay as shown in fig.\fig{ViteMedieDM}. 
Therefore, one generally expects the signals to be much below the experimental sensitivities --- with a few notable exceptions that we discuss.


\smallskip

Concerning charged currents, the leading order amplitudes are obtained from the (axial) vector interaction with right-handed leptons
\beq C_{\chi R}\big(\bar \chi \gamma_\mu e_R \big)\Big( \bar{n} \gamma^\mu \frac{1+g_A \gamma_5}{2} p \Big)\,,
 \eeq 
where $g_A \approx 1.27$~\cite{1802.01804}, and the scalar interaction with left-handed leptons $(\bar \chi  e_L) \, (\bar{n}  p)$, with a coefficient proportional to $C_{\chi L}$.
Both these interactions lead to a DM-induced $\beta$ decay. 
At the nuclear level, the relevant process is
the $\beta^-$ transition ${\rm DM} + \ce{^{$A$}_{$Z$}{\mathcal N}}  \rightarrow e^- + \ce{^{$A$}_{$Z+1$}{\mathcal N}}$, which is energetically possible if the DM mass is larger than the capture threshold $Q+m_e$, where $Q= m_{A,Z+1}-m_{A,Z}$ is the mass difference of the two nuclei. 
Following~\cite{1908.10861} the thermally 
averaged cross section arising from the charged current operator is
\beq
\langle \sigma v \rangle  = \frac{C_{\chi L(R)}^2 M^2}{16\pi} f(E_R)
\eeq
where $f(E_R)$ is a  function of the detected energy that encapsulates  the nuclear details 
of the $\beta^-$ transition. 
For light DM particles with kinetic energies much less than  $\delta = M- (m_e + Q)$, the spectrum of recoiling electrons exhibits a peak at $E_{R} \approx \delta$. 
The left panel of fig.\fig{DDID} compares the estimated sensitivities of current direct detection experiments, taken from~\cite{1908.10861},
with bounds from DM decays and from colliders.

\smallskip

One could wonder whether there is a possibility to explain the anomalous counting rate of electromagnetic recoils in the {\sc Xenon1T} detector peaked around 3 keV~\cite{2006.09721}. 
Considering Xenon as target, the process ${\rm DM} + \ce{^{$A$}_54 Xe} \to e^- \ce{^{$A$}_55 Cs}$ can have capture thresholds as low as a few hundreds keV: in particular, the $\ce{^131_54 Xe}$ isotope has an abundance of about 21\% and a capture threshold of only 355 keV~\cite{2009.00535}. One therefore can have a signal peaked at $E_R \approx 3\,{\rm keV}$ if the DM mass is $M \approx (m_e + Q)+ E_R \approx 358$~keV. 
This needs some tuning among the electron and DM masses, 
and a slightly larger $M$ 
would give a peak in the few hundreds of keV range. 
We show the best fit to the {\sc Xenon1T} data in the right panel of fig.\fig{DDID},
obtained using the rate reported in~\cite{1908.10861}. 
Assuming local DM density $\rho_\odot = 0.4\GeV/\cm^3$,
the anomaly can be fitted for values of the DM coupling $C_{\chi R}\approx 1/(3.4\TeV)^2$. 
For this value of the coupling to right-handed electrons, 
the loop-induced decay rate into $\nu_e\gamma$ gives a DM life-time longer than  the experimental bound from {\sc Planck}, as shown in the left panel of fig.\fig{DDID}. 
However, the LHC bounds on the lepton plus missing energy final state, obtained in the EFT approximation in eq.\eq{CchiRcoll}, constrain 
$C_{\chi R}$ 
to be a factor 2 smaller.\footnote{We expect $\mathcal{O}(1)$ corrections if the mediator has TeV-scale mass.
A ${\cal U}_\mu$ leptoquark mediator is exchanged in the $t$-channel,
so that bounds can become weaker, possibly making
our {\sc Xenon} interpretation compatible with the LHC bounds. In particular, the study in \cite{2011.02486} for the $(\bar{c}\gamma^\mu b)( \bar{\ell} \gamma_\mu \nu)$ operator finds $\sim$50\% weaker bounds on the new physics scale when considering the complete leptoquark model instead of the EFT description.}
A similar coupling to left-handed electrons is instead largely excluded by DM decays.
The {\sc Xenon} signal can thus most likely not be explained in our framework,
but future experiments will have the sensitivity to probe a parameter space relevant for DM freeze-in via coupling to right-handed leptons, provided that the reheating temperature is low, $T_{\rm RH} \sim \Lambda_{\rm QCD}$.

{Furthermore a novel  specific signature  from the $ ^{136}_{~55}{\rm Cs}$ decay arises. Indeed, following the analysis done in~\cite{2009.00535}  for solar neutrino capture in LXe, one could detect DM with lepton number by using a delayed coincidence signature from long-lived states of $ ^{136}_{~55}{\rm Cs}$. 

\smallskip

Concerning neutral currents, DM couplings to left-handed leptons induce
an interaction with nucleons ${\cal N}$ described at leading order by the scalar operators
\beq (\bar \chi \nu_L )( \bar{n} n)\qquad\hbox{and}\qquad (\bar \chi \nu_L )( \bar{p} p).\eeq
This is spin-independent and leads to a sort of exothermic DM-nucleus collision ${\rm DM}\, \mathcal  N \rightarrow \nu \, \mathcal  N$ with mass splitting $\delta = M- m_\nu \simeq M$. For light DM particles with kinetic energies much less than the splitting, exothermic spectra exhibit a peak at nuclear recoil energy $E_{\rm R} = \delta \, M/(2m_{\mathcal  N}) \approx M^2/(2m_{\mathcal  N})$, where $m_{\mathcal  N}$ is the mass of the target nucleus.
Hence, for Xenon target we expect detectable nuclear recoil in the keV range when the DM mass is around 20 MeV.
Nuclear recoil affects atomic electrons, inducing extra bremsstrahlung and Migdal emission.
It is worth noticing that the window of DM masses probed in such scenario is much smaller than the one achieved with standard DM-nucleus collisions. However, the spectrum is analogous to the one of  exothermic collisions and therefore  we cannot discriminate further among them.

\begin{figure}[t]
\centering%
\includegraphics[width=0.48\textwidth]{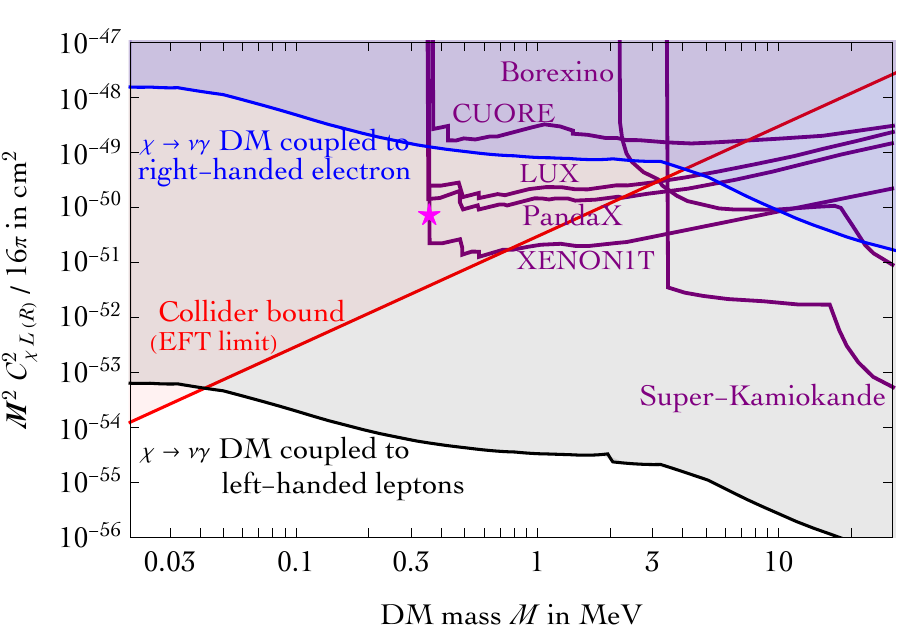}\hfill
\includegraphics[width=0.48\textwidth]{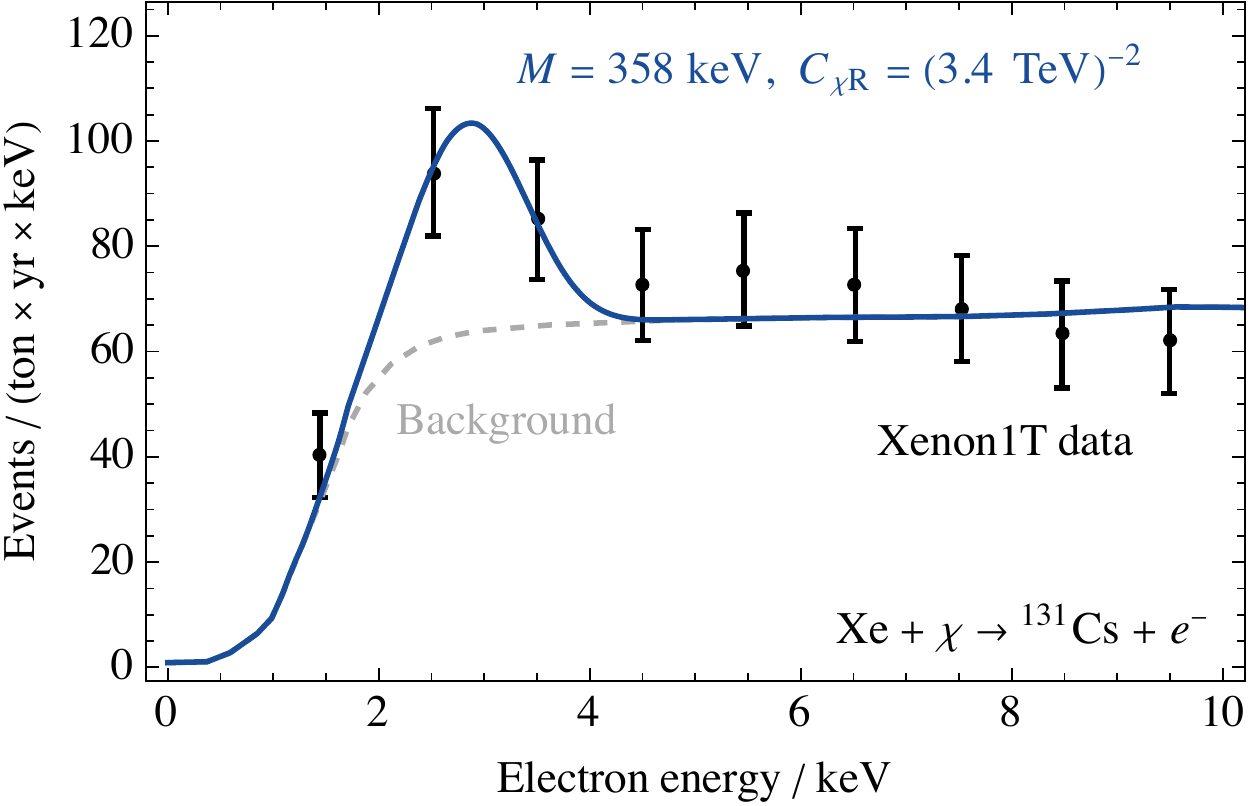}
\caption{{\bf Left:} Estimated experimental sensitivities for the DM absorption cross section (from~\cite{1908.10861}), 
compared to the indirect detection bound on DM life-time
(these are stronger than in~\cite{1908.10861}, where
the dominant QCD contribution to $\DM \to\nu\gamma$ was not considered)
and to LHC bounds estimated in the EFT limit (red line).
The magenta star shows the parameters that can fit the {\sc Xenon1T} excess. 
{\bf Right:} Best fit to the {\sc Xenon1T} electron-recoil data from DM absorption on $\ce{^131 \!Xe}$.
\label{fig:DDID}}
\end{figure}

%
%

\medskip

DM absorption signals have been studied previously in~\cite{1908.10861}.
However, when discussing neutral currents these authors considered a $(\bar\chi \nu_R)(\bar q q)$ effective
interaction involving $\nu_R$ (rather than $\nu_L$ or $\ell$)
in order to avoid too fast DM decay $\chi \to \nu_R\gamma$.
Since $\nu_R$ has no SM gauge interactions their signals are equivalent to those of inelastic DM with 
two states $\chi$, $\chi'$ and a $(\bar\chi \chi') (\bar q q)$ interaction.
When discussing charged currents \cite{1908.10861} considers our same $(\bar \chi \ell)(\bar d u)$ interaction,
but does not include the DM decay channel $\chi \to \nu\gamma$ mediated by QCD states at one loop.
We report in the left panel of fig.\fig{DDID} the projected constraints for the DM-induced $\beta^-$ signals taken from~\cite{1908.10861}. One can see that, while DM coupled to right-handed electrons can give observable signals in various experiments, for DM coupled to left-handed leptons the signals compatible with DM life-time are always below the experimental sensitivity.
%
%
%

%
%
%
%
%
%

%
%
%
%
%

\subsection{Indirect detection signals and the 3.5 keV anomaly}\label{ID}
DM decays provide indirect detection signals:
the constraints on
tree-level and loop-induced DM decay from {\sc Planck} and {\sc Voyager}
have already been discussed in section~\ref{ellpidecays}.
As these constraints can be saturated, signals from DM decays are detectable.

\smallskip


In particular, the  unidentified $X$-ray emission line claimed at 3.5 keV~\cite{1402.2301,1402.4119} 
could be due to DM with mass $M \approx 7\keV$ decaying as $\DM\to\nu_\ell \gamma $
with rate $\Gamma \approx 1.4 ~ 10^{-28}/\s$ (see e.g.\  fig.~13 of \cite{1402.2301}).
In the considered models the needed DM life-time is obtained for
\beq C_{\chi L} \approx 10^{-4} \frac{4\GF}{\sqrt{2}}
\qquad\hbox{or}\qquad
C_{\chi R} \approx \frac{4\GF}{\sqrt{2}}
\left\{\begin{array}{ll}
6~ 10^{-2} & \hbox{DM coupled to $\ell=e$}\\
3~ 10^{-4} & \hbox{DM coupled to $\ell=\mu$}\\
2~10^{-5}& \hbox{DM coupled to $\ell=\tau$}
\end{array}\right. .\eeq
For such values of the couplings
the relic DM abundance can be reproduced via the freeze-in  mechanism with a low reheating temperature $T_{\rm RH} \sim \Lambda_{\rm QCD}$, see fig.~\ref{fig:DMmupiSummary}.
This is a non-trivial result, 
as other models where DM is a sterile neutrino can fit the 3.5 keV line, 
but its relic abundance cannot be obtained via freeze-in as non-resonant  oscillations:
extra mechanisms  are needed,
such as decays $S\to \DM\,\DM$ of an ad-hoc scalar $S$~\cite{hep-ph/0609081} (possibly with a partially closed phase space, if DM needs to be colder~\cite{Heeck:2017xbu}).
In our model on the other hand the leptoquark motivated by flavour anomalies plays a similar role,
decaying into one DM particle and one SM particle.

%

%
%

Concerning scatterings  that could induce a gamma-ray signal, the only 
relevant interactions are those that  change the DM number by one. Indeed 
  the pair-annihilation of DM particles is extremely suppressed within the freeze-in production mechanism,
  as $\langle \sigma_{\rm ann} v \rangle \propto g_\chi^4$ and the $g_\chi$ coupling is very small.  
  
\smallskip  
  
We thereby focus on further indirect detection gamma-rays signals coming from DM decay and from 
scatterings that change DM number by one unit.
A possible way to get a signal is to change the DM number by one  via the scatterings with highly energetic  Galactic Cosmic-Ray (CR) electrons and protons.
The relevant primary channels are $\chi e^+ \to \bar d u$ and $\chi \bar p \to \bar n e$ and the corresponding charge conjugates. 
The spatial morphology of the induced gamma-ray signal is non-conventional and is enhanced in regions with high density of baryons and DM. 
As a consequence dwarf galaxies, which are usually considered as the cleanest laboratories in indirect detection, 
are not the best place to look for gamma-rays signals from DM with lepton number.  
The energy spectrum is similar
to the one  from standard pair-annihilation with the difference that the total energy is not $4 M^2$  but $\left(M + E_{\rm CR}\right)^2$
with $E_{\rm CR}$ the  energy of Galactic CR. 
The cross section of highly energetic Galactic CR electrons and protons
scattering off non-relativistic DM is
\beq
 \sigma \approx  {C_{\chi L(R)}^2} \left(M + E_{e, p}\right)^2 = 
5.3 \times 10^{-10}\,  {\rm pb} \left(\frac{C_{\chi L(R)}}{10^{-4}\GF}\right)^2 \left(\frac{M + E_{e, p}}{\rm{GeV}}\right)^2 \ .
 \eeq 
The highest $C_{\chi L(R)} \sim 10^{-4} \GF$ allowed by DM stability and consistent with the freeze-in mechanism with  low reheating temperature $T_{\rm RH} \sim \Lambda_{\rm QCD}$ 
is  obtained for very light DM. Hence we predict $\sigma \lesssim 10^{-9}  {\rm pb} \, (E_{e, p}/{\rm GeV})^2$, which gives a flux too low to be detected by any indirect detection experiments.

\section{Flavour models with multiple DM and freeze-out}\label{DM2}

So far we considered minimal models with one DM particle, and we could not
reproduce its cosmological abundance from freeze-out via thermal decays,
compatibly with bounds on the DM lifetime, $\tau \circa{>} 10^{26}$~s, although we found viable models assuming DM freeze-in.

We here show how bounds on the DM lifetime can be easily satisfied in the presence
of two or more DM states $\chi_{i}$.\footnote{This is to be contrasted to DM models where a cosmologically short-lived state freezes out due to pair annihilations, and later decays to one or more SM states plus a cosmologically long-lived (so-called super-WIMP) DM state, 
discussed e.g.\ in~\cite{hep-ph/9905212,hep-ph/0306024}.
In this case DM decays are slower than the Hubble time, and play no role in fixing the initial DM freeze-out abundance:
they only enter later reducing it by the ratio of the two masses  in the dark sector.
This  `freeze-out and decay' scenario can be extended to freeze-out via thermal decays, but it contains
out-of-equilibrium elements. We prefer to focus on strict freeze-out.}
This extension could be motivated by flavour: for example,
in the models with extra vectors such as eq.\eq{LQ} 
one would naturally expect 3 generations of DM singlets $\chi_{1,2,3}$.

To keep the discussion simple and general, we summarize 
our previous results about the decay rate of one DM particle as
\beq \Gamma_{\rm DM} \sim\epsilon_{\rm SM} \epsilon_{\rm DM} M \qquad
\hbox{where} \qquad \epsilon_{\rm DM}\sim C^2 M^4,\qquad
 \quad \epsilon_{\rm SM}\sim \frac{\GF ^2 M^4}{(4\pi)^4}.
\eeq
$ \epsilon_{\rm SM}\sim 10^{-18}$ is a typical electroweak loop factor at the DM mass $M\circa{<}\GeV$.
If $C \sim \GF$, as needed to have thermal freeze-out at $M \sim \GeV$,
the two $\epsilon$ suppression do not give a stable enough DM
(while freeze-in needs $C \ll \GF$, improving DM stability).

Stable enough DM  is obtained in the presence of two states, $\chi_1$ and $\chi_2$,
with mass $M_{1,2}\circa{<}\GeV$ (arising, for example, as $m_\ell + m_\pi$) 
and effective interactions of the type
\beq C_{\rm DM_1}(\bar \chi_1 \ell )(\bar d u)+
C_{\rm DM_2}(\bar \chi_2 \ell )(\bar d u)+
C_{\rm DM_{12}}(\bar \chi_1 \chi_2 )(\bar q q).
\eeq
Notice that we have not imposed any $\mathbb{Z}_2$ symmetry that keeps the dark sector separated from the SM sector.
We next assume that $C_{\rm DM_1}$ is negligibly small, 
and that masses $M_1<M_2$ are in the range such that
$\chi_2 \to \ell \pi$ decays  are kinematically  closed, while
$\chi_2 \to \chi_1 \pi^0$ decays are kinematically open
(we make this latter assumption just for simplicity;
we could relax it and use $\chi_2 \to \chi_1 \gamma\gamma$ decays).
Then the decay widths are
\beq
 \Gamma_{\rm DM_2} \sim  \epsilon_{\rm DM_{2}}M_2,\qquad
  \Gamma_{\rm DM_1}\sim  \epsilon_{\rm DM_{12}} \frac{ \Gamma^*_{\rm DM_2}}{M_2}  M_1
  \sim
 \epsilon_{\rm SM} \epsilon_{\rm DM_{2}}  \epsilon_{\rm DM_{12}}M_1\eeq
 where $  \epsilon_{\rm DM_{12}} \sim C^2_{\rm DM_{12}}M_{1,2}^4$. 
The smallness of  $C_{\rm DM_1}$, at a level that saturates the decay width $  \Gamma_{\rm DM_1}$ above, could be motivated by flavour considerations (for instance a very precise flavour alignment in the right-handed lepton sector, if $\chi$ carry lepton flavour number).
Quantum effects respect this requirement.
Assuming now that $C_{\rm DM_2}\sim C_{\rm DM_{12}} \sim \GF$ or even larger depending on the masses, we have, at the same time:
\begin{enumerate}
\item Very long lived $\chi_1$ DM, that satisfies bounds on the DM life-time;

\item Cosmologically fast decays of $\chi_2$, so that it does not pose cosmological problems;

\item  Thermal freeze-out via $\chi_1 \leftrightarrow \chi_2 \leftrightarrow {\rm SM}$ 
scatterings at finite temperature.

\end{enumerate}
Although not necessary, DM stability can be improved by an extra $\epsilon_{{\rm DM}_{23}}$ factor 
adding a third generation $\chi_3$ and repeating the above structure:
each mediation step can give one extra $\epsilon$ suppression
at zero temperature, while rates at finite temperature can reach thermal equilibrium, $\Gamma\sim H$
at $T\sim M$,
if $C_{{\rm DM}}\sim \GF $.
This latter condition (needed to have thermal equilibrium down to the DM mass) can be borderline or
problematic in specific models.

\section{Conclusions}\label{concl}
We explored the possibility of matching the cosmological DM density via 
particle-physics processes that change the total DM number (DM particles plus their anti-particles)
by one unit, dubbed {\em thermal decays}.
Differently from the usual models where two DM particles annihilate,
no $\mathbb{Z}_2$ symmetry is introduced to distinguish the dark sector from the SM sector,
and to keep DM stable.

In section~\ref{Boltzmann} we wrote the relevant Boltzmann equations and provided approximate analytic solutions, finding
that thermal decays can be much more efficient than DM annihilations,
and thereby can need much lower rates.
In the most favourable situation, thermal decays just need to reach thermal equilibrium at temperatures around the DM mass.
Fig.\fig{NeededgG} shows the needed values of DM couplings:
dimension-less couplings need to be larger than $10^{-9}$, and dimension-6 operators
can be suppressed by a scale only mildly larger than the weak scale.

\medskip

The price is that DM can decay, and the bounds on its life-time are 30 orders of magnitude stronger than the
DM rates needed for thermal freeze-out.
So the above possibility seems largely excluded.
To avoid this conclusion
we considered DM in co-stability ranges such that main DM decays at zero temperature are kinematically blocked,
while scatterings at finite temperature are open.
Kinematical blocking allows to improve DM stability by one electroweak loop factor --- about 18 orders of magnitude 
if the DM mass is around the muon or pion masses.
This is not enough to allow for DM thermal freeze-out, and it is enough to allow DM freeze-in.

In section~\ref{DM1} we implemented this mechanism in models where DM is
a right-handed neutrino: a SM singlet with lepton number one.
Such a state can interact with SM fermions in presence of an extended gauge group, or additional generic mediators.
%
%
This includes, for example, the TeV-scale Pati-Salam leptoquark recently considered to fit the $B$-physics anomalies.
We find that these models can successfully reproduce the DM abundance via freeze-in,
with small DM couplings possibly compatible with a Froggat-Nielsen-like picture of flavour.

We discussed possible specific signatures of these models, depending on the DM mass. Concerning indirect detection, DM decays
for example allow to fit the 3.5 keV anomaly compatibly with freeze-in DM production,
if $M\approx 7\keV$ and the reheating temperature is low. 
Concerning direct detection, one has unusual processes where DM can convert into SM particles.
An explanation of the {\sc Xenon1T} excess peaked around 3 keV
is compatible with bounds from DM decays and freeze-in production,
but disfavoured by LHC bounds. A dedicated experimental search is therefore needed to precisely establish order one factors.
The collider signals are those of heavy vectors, and flavour signals include their effects.

In section~\ref{DM2} we found that extra suppression of DM decays, as needed to 
reproduce the DM cosmological abundance via freeze-out of thermal decays,
can be obtained in models with two or more new states.
Their presence could be motivated, for example, by flavour considerations.

\small

\paragraph{Acknowledgements}
We thank Eugenio Del Nobile, Jeff A.~Dror, Elena Graverini, Darius Faroughy and Tomer Volansky for discussions.
This work was supported by MIUR under contracts PRIN 2017L5W2PT and PRIN 2017FMJFMW, and by the ERC grant 669668 NEO-NAT.

\appendix

\section{Rates in the model with DM coupled to $\ell \pi$}\label{appA}
We start summarizing the pion action~\cite{hep-ph/0505265}. We define
\beq U = \exp \frac{ i \Pi}{f_\pi}\,, \qquad
\Pi = \begin{pmatrix}
\pi^0 & \sqrt{2} \pi^+ \cr \sqrt{2}\pi^- &- \pi^0
\end{pmatrix} \,, \eeq
where $f_\pi =93\MeV$. 
The SM pion Lagrangian is given by 
\beq \label{eq:SMpi}\Lag_{{\rm SM}\pi} = \frac{f_\pi^2}{4} \Tr \left [|D_\mu U|^2 + m_\pi^2 (U + U^\dagger)\right] \,, \eeq
where the covariant derivative is  
\beq
  D_\mu U = \partial_\mu U - i (r_\mu^A + r_\mu^W) U + i U (l_\mu^A + l_\mu^W)  \,, 
\eeq
with 
\beq
r_\mu^A = l_\mu^A = -e Q A_\mu\,, \qquad  
 r_\mu^W = 0\,,  \qquad
 l_\mu^W = -{g V_{ud} \over \sqrt 2} (\sigma^+ W_\mu^+ + \sigma^- W_\mu^-) \eeq
where 
 \beq  Q = {\sigma^3 \over 2} = {1\over2} \begin{pmatrix} 1 & 0 \\ 0 & -1 \end{pmatrix}\,,
 \qquad \sigma^+ = \begin{pmatrix} 0 & 1 \\ 0 & 0 \end{pmatrix}\,, \qquad \sigma^- = \begin{pmatrix} 0 & 0 \\ 1 & 0 \end{pmatrix}\,. 
\eeq
The Lagrangian is then expanded as
\begin{eqnarray}\nonumber
\Lag_{{\rm SM}\pi} &=& \frac12 (\partial_\mu \pi^0)^2 + |\partial_\mu \pi^+|^2  - m_\pi^2\left(|\pi^+|^2 + \frac12  (\pi^0)^2\right)
   \\ \nonumber
&&+ \Big( -\frac{g }{2} f_{\pi} V_{u d} W_{\mu}^+ (\partial^\mu - i e A^{\mu}) \pi^- + i e A_{\mu} \pi^+ (\partial^\mu - i e A^{\mu}) \pi^- 
\\ \nonumber
&&~- \frac{i g }{2}V_{u d} W_{\mu}^+ \left[ \pi^0 (\partial^\mu - i e A^{\mu}) \pi^- - \pi^- \partial^\mu \pi^0 \right]+ \textrm{h.c.} \Big) +
\\
&&+ \frac{1}{6f_\pi^2} \bigg[\big((\pi^-\partial_\mu \pi^+)^2  +2 \pi^0\pi^- \partial_\mu \pi^+\partial^\mu \pi^0  + \textrm{h.c.} \big)+
  \nonumber
\\
 \nonumber
&&~- 2 |\partial_\mu \pi^+|^2 (|\pi^+|^2 + (\pi^0)^2) - 2|\pi^+|^2 (\partial_\mu\pi^0)^2 + m_\pi^2\left(|\pi^+|^2 + \frac12 (\pi^0)^2\right)^2
\bigg] +\cdots
\end{eqnarray}
where we only wrote the relevant terms in the expansion.
Integrating out the $W$ gives the effective Fermi interaction of pions
\beq
  \Lag_{{\rm SM}\pi}^{\rm weak} =  2 \GF   (\bar\nu_L\gamma^\mu \ell_L) \big( f_\pi \partial_\mu\pi^+  - i  (\pi^0\partial_\mu\pi^+ - \pi^+\partial_\mu\pi^0) + \ldots \big) + \textrm{h.c.} 
\eeq
The vector-like DM interaction in eq.\eq{LagDMellpi} is similarly obtained, replacing the left-handed current with $(\bar\chi \gamma^\mu \ell_R)$.
The scalar interaction is obtained converting $ (\bar\chi L_L)(Q_L u_R)$  into
\beq
 (\bar \chi \nu_L) \frac{f_\pi^2 B_0}{2}{\rm Tr}\Big[\frac{\sigma^3}{2}U + \frac{1}{2}U\Big] +  (\bar\chi \ell_L)\frac{f_\pi^2 B_0}{2}{\rm Tr}\Big[\sigma^+ U\Big] .
\eeq

\subsection{Tree-level DM decay}\label{DMellpitreedecay}
 If $M  > m_\ell$, the DM can decay to a lepton and an off-shell pion (subsequently decaying to $e\bar{\nu}$) through the interaction in eq.\eq{LagDMellpi}. 
 The amplitude reads
 \begin{equation}
 \mathscr{A} = 
 \sqrt{2} C_{\chi R}  \GF  
 \frac{i}{p_\pi^2 - m_\pi^2}f_\pi^2 p_\pi^\mu (-p_\pi^\nu) \left[ (\bar{u}_\chi \gamma_\mu P_R v_\ell)(\bar{u}_e\gamma_\nu P_L v_\nu) + (\bar{u}_\ell \gamma_\mu P_R v_\chi)(\bar{u}_e\gamma_\nu P_L v_\nu) \right],
 \end{equation}
 where the second term is absent if $\chi$ is Dirac. The squared amplitude is
 \begin{equation}
 \overline{|\mathscr{A}|^2} = 
 (2)\frac{2 f_\pi^4 \GF ^2 C_{\chi R}^2 
 m_e^2 (p_\pi^2 - m_e^2)}{(m_\pi^2 - p_\pi^2)^2} \left[(M ^2 - m_\ell^2)^2 - p_\pi^2(M ^2 + m_\ell^2)\right]\,
 \end{equation}
 where $p_\pi = p_e + p_\nu$ is the pion 4-momentum and the (2) holds only if $\chi$ is Majorana.
 The decay width is
 \begin{align}
 \Gamma(\chi\to \ell e\nu) &= \frac{1}{2M }\int_{s_{12}^{\rm min}}^{s_{12}^{\rm max}}\int_{s_{23}^{\rm min}}^{s_{23}^{\rm max}} ds_{12} ds_{23} |\mathscr{A}|^2 \frac{1}{128\pi^3 M ^2}\\ 
&= \frac{1}{2 M } \int_{m_e^2}^{(M  - m_\ell)^2} \frac{dp_\pi^2}{128\pi^3 M ^2} \, |\mathscr{A}|^2 \left(1 - \frac{m_e^2}{p_\pi^2}\right)\sqrt{m_\ell^4 - 2 m_\ell^2(M^2 + p_\pi^2) + (M ^2 - p_\pi^2)^2}.
 \end{align}
Off-shell decay rates have been computed both analytically and numerically with {\sc MadGraph}~\cite{1405.0301}, after having implemented the interaction in eq.\eq{LagDMellpi} and the relevant SM pion interactions.
 
\subsection{Scatterings among $\chi,\ell,\gamma, \pi^\pm$}\label{extragamma}
The photon interacts with charged particles, $\pi$ and $\ell$. 
Furthermore a $\chi \ell \gamma \pi$ vertex arises from the gauge-covariantization 
of the $\chi \ell \partial_\mu\pi$ interaction.
Taking this vertex into account, the total scattering amplitude vanishes for longitudinal photons.
In the non-relativistic limits 
the squared amplitudes summed over all initial-state and final states are
\begin{eqnsystem}{sys:AA}
 |\mathscr{A}|^2 (\chi \gamma \to \pi^+ \ell) &=& \frac{e^2 f_\pi^2 C_{\chi R}^2 (m_\ell+m_\pi)(m_\ell^2+M^2)}{m_\ell}, \\
 |\mathscr{A}|^2 (\chi \pi^- \to \gamma \ell) &=& \frac{e^2 f_\pi^2 C_{\chi R}^2 M(m_\ell^2+M^2)}{m_\pi+M}, \\
 |\mathscr{A}|^2 (\chi \ell \to \pi^-\gamma ) &=& \frac{e^2 f_\pi^2 C_{\chi R}^2 M(m_\ell^2+M^2)}{m_\ell},
\end{eqnsystem}
with no contribution from the photon interaction to the $\pi$.

\subsection{Scatterings among $\chi,\ell,\pi^0, \pi^\pm$}\label{extrapi}
Only the quartic interaction among $\chi,\ell_R,\pi^0, \pi^\pm$ contributes to tree-level scatterings among these particles.
All processes with different initial-state and final-state particles have
the following squared amplitude, summed over all degrees of freedom, 
and here written assuming all incoming momenta, $p_\chi+p_\ell +p_{\pi^0}+p_{\pi^\pm}=0$:
\begin{eqnarray} 
  |\mathscr{A}|^2 = -2 C_{\chi R}^2 \left\{
  \left[ p_\ell \cdot ( p_{\pi^0} - p_{\pi^+} ) \right]  \left[ p_\chi \cdot (p_{\pi^0} - p_{\pi^+} ) \right]
+ \left( p_{\pi^+} \cdot p_{\pi^0} - m_{\pi}^2  \right) (p_\ell \cdot p_\chi)\right\} \, .
\end{eqnarray} 
 So the  squared amplitudes, summed over all degrees of freedom, are
\begin{eqnarray*} 
 |\mathscr{A}|^2_{\chi \ell \to \pi^- \pi^0} &=& 
  -\frac{f_\pi^2 C_{\chi R}^2}{2} 
 [4m_\pi^4+4t^2+(m_\ell^2+M^2)(m_\ell^2+M^2-s)+4t(s-m_\ell^2-M^2-2m_\pi^2)],\\
  |\mathscr{A}|^2_{\chi \pi^+ \to \bar\ell \pi^0}  &=&
  -\frac{f_\pi^2 C_{\chi R}^2}{2} 
  [4m_\pi^4+m_\ell^4+M^4 + m_\ell^2(2M^2-4s-t)+4s(s+t-2 m_\pi^2)-M^2(4s+t)],
 \end{eqnarray*} 
and the same for  $\chi \pi^0 \to \bar\ell \pi^-$. 

For each scattering, either the initial state or the final state can be non-relativistic, depending on particle masses.
The $s$-wave non-relativistic cross section for a generic $1+2\to 3+4$ scattering with 
$m_1+m_2 > m_3 +m_4$ (so that the initial state is non-relativistic) is
\beq
\sigma_0 \equiv \sigma v_{\rm rel}  \simeq \frac{|\mathscr{A}|^2}{32\pi m_1 m_2 d_1 d_2}
\sqrt{ \bigg[ 1 - \bigg(\frac{m_3+m_4}{m_1+m_2}\bigg)^2\bigg]
\bigg[ 1 - \bigg(\frac{m_3-m_4}{m_1+m_2}\bigg)^2\bigg]}
.\eeq

\end{document}